\documentclass[aps,prl,reprint,superscriptaddress,longbibliography]{revtex4-1}
\usepackage{amsmath}
\usepackage{amsfonts}
\usepackage{color}
\def \ba {\begin{eqnarray}}
\def \ea {\end{eqnarray}}

\usepackage{epsf,latexsym,graphicx}
\usepackage{xr}
\usepackage{xcite}
\externalcitedocument{Crossover_supp}

\begin{document}

\title{Dynamical effects on superconductivity in BCS-BEC crossover}

\author{Tae-Ho Park}
\affiliation{Department of Physics and Institute for Basic Science Research, Sungkyunkwan University, Suwon 16419,
Korea}

\author{Han-Yong Choi}
\affiliation{Department of Physics and Institute for Basic Science Research, Sungkyunkwan University, Suwon 16419,
Korea}
\affiliation{Asia Pacific Center for Theoretical Physics, Pohang 37673, Korea}

\date{\today}

\begin{abstract}

We investigate the dynamical effects of pairing interaction on superconductivity in BCS-BEC crossover by studying the Holstein model at half-filling where the electron-phonon coupling $g$ controls the crossover. The dynamical mean-field theory was employed in combination with the numerical renormalization group technique. The dynamical effects induce distinct features such as absence of the dispersion back-bending of Bogoliubov quasi-particles, non-monotonous coupling dependence of the pairing gap, and the soft phonon spectrum due to the Goldstone mode of local pair phase fluctuations. Also interesting is the maximum critical temperature being at the normal state phase boundary.
Some of these features have intriguing similarities with the recent observations in the FeSe$_{1-x}$S$_x$ and Fe$_{1+y}$Se$_x$Te$_{1-x}$ iron-chalcogenides in the BCS-BEC crossover.

\end{abstract}

\pacs{}
\maketitle

$Introduction$\label{sec:intro}
--
It suits nature of physics to envisage a unified theory to understand the Bardeen-Cooper-Schrieffer (BCS) theory of superconductivity and Bose-Einstein-condensation (BEC) of composite bosons on an equal footing \cite{griffin_1994,leggett_2012,randeria_2014}. This BCS-BEC crossover was investigated by Eagles in late 60's \cite{eagles_1969} and later by Leggett \cite{leggett_1980} and others \cite{nozieres_1985,randeria_1989,melo_1993,engelbrecht_1997,chen_2005} by employing the attractive-Hubbard-like models in continuum or discrete lattices.
An important conclusion is that the change between the BCS and BEC regimes is gradual so that it is permissible to think of them as two facets of a unified theory. More concretely, the standard picture of the BCS-BEC crossover from the attractive-Hubbard-like models is that the pairing amplitude $\Delta_p$ increases monotonically while the superconducting (SC) critical temperature $T_c$ increases to have a maximum and decreases as one goes from the BCS to BEC regimes.
The change between the regimes can be tuned by varying the ratio $\Delta_p/\epsilon_F$, where $\epsilon_F$ is the Fermi energy, or the strength of the pairing interaction.

Experimental investigations into the crossover have not progressed much because of the difficulties in realizing quantum materials in the BEC regime of $\Delta_p/\epsilon_F \sim 1$, whereas the ratio is $ \sim 10^{-4}$ for conventional superconductors.
Then came a burst of research activities because cuprate high-temperature superconductors in the underdoped pseudogap region seem to be in the BEC regime ($\Delta_p/\epsilon_F\sim1$) where the carrier density and $\epsilon_F$ are strongly suppressed by the Mott physics \cite{Lee_2006}.
However, the cuprate pseudogap seems to exhibit phenomena too much complex to be understood in terms of the BCS-BEC crossover \cite{randeria_2014}.

This crossover has also been extensively studied in dilute ultracold Fermi alkali gases by tuning the interaction between the Fermi atoms using a Feshbach resonance \cite{gaebler_2010,regal_2005,bloch_2008,chin_2010}. This has revealed many insights into the unitary limit which lies in between the BCS and BEC regimes.

Recently, interest in the BCS-BEC crossover in SC materials has been renewed in iron-chalcogenide superconductors because it apparently seems that they can be tuned into both regimes. FeSe doped with isovalent S, FeSe$_{1-x}$S$_x$, is in the nematic phase for $x \lesssim 0.17$ and in the tetragonal phase for $x \gtrsim 0.17$. The specific heat $C_p$  measurements exhibit the classic BCS behavior with a finite jump at $T=T_c$ at $x=0$ and 0.1, but, at $x=0.2$, show $C_p/T \propto\sqrt{T/ T_c}$ below $T_c$ and the archetypical BEC behavior of the $\lambda$ transition \cite{sato_2018,mizukami_2018}. $T_c$ shows a broad maximum and decreases as $x$ changes between 0 and 0.25 \cite{sato_2018,mizukami_2018}. However, the scanning tunneling microscopy measurements show that $\Delta_p$ decreases as $x$ is increased into the BEC regime contrary to the standard BCS-BEC crossover picture \cite{hanaguri_2018}.

Also interesting is the angle-resolved photoemission spectroscopy study on Fe$_{1+y}$Se$_x$Te$_{1-x}$ by Kanigel group \cite{rinott_2017,lubashevsky_2012}. They altered the Fe content $y$ for a fixed $x=0.4$ to tune the chemical potential, and measured the SC quasi-particle dispersion as the system evolves from BCS to BEC regimes. They found that as $\Delta_p/\epsilon_F $ increases, the dispersion changes from a characteristic back-bending near $k_F$ in BCS regime to a minimum gap at $k=0$ in BEC regime, consistent with the theoretical results of Loh et al \cite{Loh_2016}. Our study to be presented here indicates this picture needs to be elaborated on.

Now that the crossover is being explored experimentally in the condensed materials and that it is the frequency dependence of the interaction which underlies pairing at least for the conventional superconductors, it seems worthwhile to investigate the frequency dependence of pairing interaction in the BCS-BEC crossover. In this Letter, we build on this idea and show that there are dynamically induced interesting features in the BCS-BEC crossover such as absence of the dispersion back-bending, non-monotonous coupling dependence of the pairing gap, and the maximum $T_c$ at the critical coupling concomitantly with the emergence of the Goldstone mode.
We will focus on superconductivity in the Holstein model of electron-phonon coupled systems defined in Eq.\ (1) below to explicitly present these features.
They seem to have intriguing similarities with the recent observations in the iron-chalcogenides in the BCS-BEC crossover, yet, our goal is to gain new insights into the BCS-BEC crossover brought in by the dynamical effects rather than to describe specific experiments.

$Model $\label{sec:model}
--
The Holstein model is given by
\begin{equation}\label{eq:Hol}
{\cal H} = -\frac{t}{\sqrt{q}} \sum_{\langle i,j\rangle\sigma}
c_{i\sigma}^\dag
c_{j\sigma} +\omega_0 \sum_i a_i^\dag a_i
+ g \sum_i ( a_i^\dag +a_i) \left( n_i -1 \right),
\end{equation}
where $c_{i\sigma}$ and $a_i $ are the fermion of spin $\sigma$ and boson operators at the site $i$, and $\langle i,j\rangle$ implies the nearest neighbors with the coordination number $q$, and
$n_i =\sum_\sigma c_{i\sigma}^\dag c_{i\sigma}$ is the electron density operator at the site $i$. The electrons are coupled with the Einstein
phonon of frequency $\omega_0$ with the onsite coupling constant $g$.
The model has been explored previously via various tools including the dynamical mean-field theory (DMFT) technique \cite{benedetti_1998,meyer_2002,capone_2003,jeon_2004,koller_2004}.

A useful way to get a grip on this model to integrate the phonons out and write the effective interaction between electrons as
 \ba
 U_{eff} (\omega) = \frac{2g^2 \omega_0}{\omega^2 -\omega_0^2 }.
 \ea
In the limit $\omega_0 \rightarrow \infty $, $ U_{eff} (\omega)\rightarrow -E_b$, where $E_b =2g^2/\omega_0$ is the bipoalron energy.
The Holstein model in the regime $\omega_0 \gg D $ is mapped onto the attractive-Hubbard model with the onsite interaction $|U|=E_b$ and the $\omega$ dependence may be neglected. $D=2t$ is the half-bandwidth and is equal to $\epsilon_F$ for half-filled cases.
Then, the dynamical effects become pronounced for adiabatic cases of $\omega_0 \ll D$.  This regime, however, has been a challenge to treat theoretically because of the different energy scales of electrons and phonons and of emergent superconductivity and soft boson modes \cite{deppeler_2002}. The Holstein model with superconductivity has been calculated previously only for anti-adiabatic regime using DMFT with the continuous-time quantum Monte Carlo technique \cite{murakami_2013,murakami_2014}.

Here, we solve the Holstein model at half-filling ($\langle n\rangle=1$) at zero temperature employing DMFT in combination with Wilson's numerical renormalization group (NRG) technique. The NRG technique can zoom in, via a logarithmic discretization of the conduction band, onto the low energy with an arbitrary precision and is ideal to address the small energy scales in this problem \cite{bauer_2009,bulla_2008,jeon_2003,hewson_2002}. We will consider $\omega_0=0.1$ and $\omega_0=2$ as representative cases for adiabatic and anti-adiabatic phonons. We take $D$ as the unit of energy in this paper. Details of the formulation and calculations are given in the Supplementary Material.

\begin{figure}
\begin{center}
\includegraphics[scale=0.65]{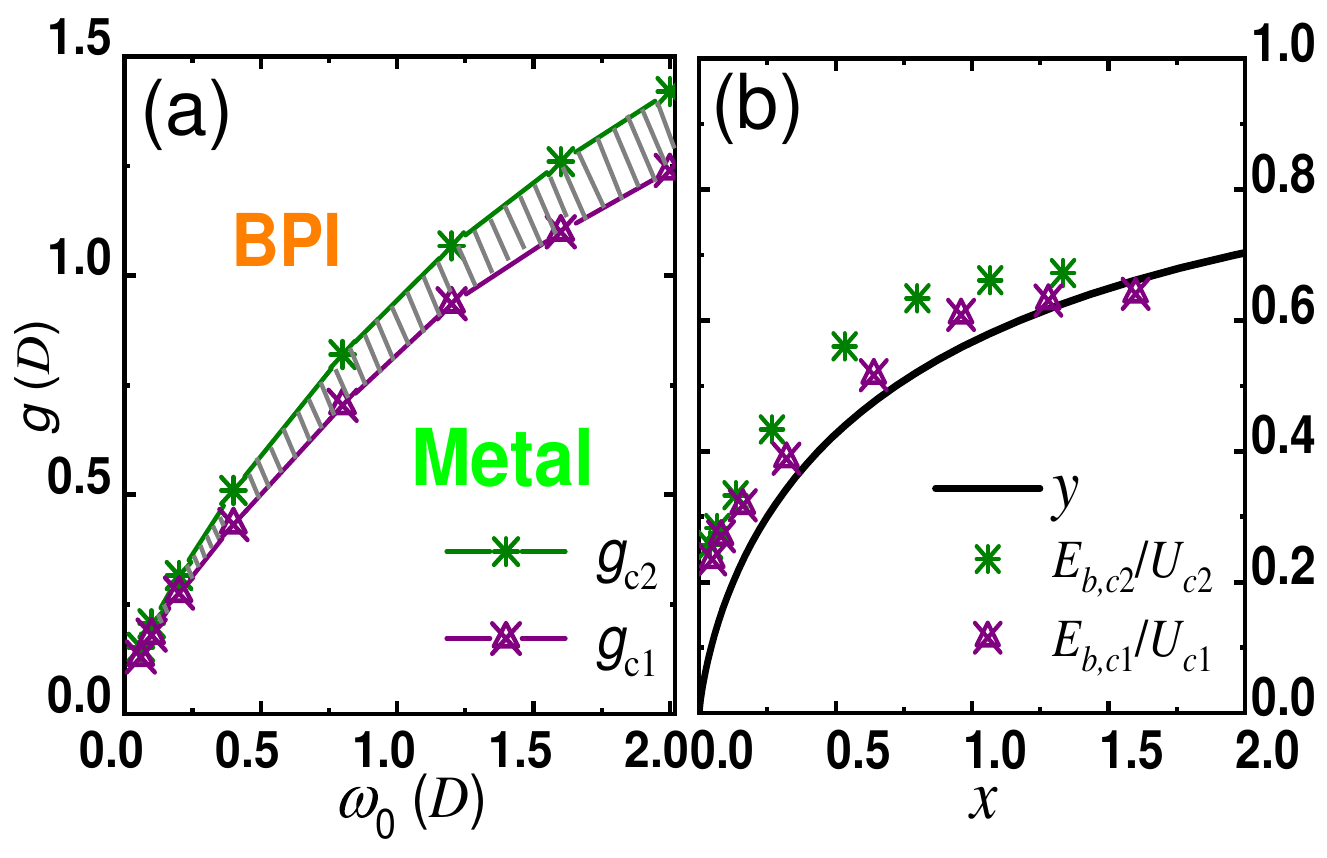}
 \caption{(a) The normal state phase diagram of the Holstein model in the plane of $\omega_0$ and $g$. The unit of energy is $D$. The ground state is either metal or BPI, where BPI stands for the bipolaron insulating phase. The shaded area is the coexisting region of the metal and BPI. (b) Plot of $g_{c1}$ and $g_{c2}$ vs.\ $\omega_0$ of (a) against the scaling function $y/\ln(y) = -x$ to exhibit the relevance of the polaronic band narrowing effects. See the text for more details. }
\label{fig:phased_Hol}
\end{center}
\end{figure}

{\it Normal state }\label{sec:resultsA}
--
Let us recall what has been established in the normal state \cite{paci_2005,capone_2006,yun_2007,barone_2008}.
In the absence of long range orders, the model exhibits the 1st order metal-insulator transition as a function of $g$. The insulating phase is a bipolaron insulator (BPI), that is, a local pair state. Like the repulsive or attractive-Hubbard model, there exist two critical values $g_{c1}$ and $g_{c2}$, such that for $g<g_{c1}$ the ground state is a metallic state, for $g>g_{c2}$ a BPI state, and for $g_{c1} \leq g \leq g_{c2}$, both metallic and insulating solutions coexist.
We plot in Fig.\ \ref{fig:phased_Hol} the $g_{c1}$ and $g_{c2}$ as a function of $\omega_0$.

As was reported previously,\cite{yun_2007}
$g_{c1}$ and $g_{c2}$ may be estimated by noticing that (a) $U_{c1}/D \approx 2.5$ and $U_{c2}/D \approx 3.0$ for the Hubbard model, and (b) the electron bandwidth is narrowed such that $D \rightarrow D e^{-(g/\omega_0)^2} $ because of the polaronic effects \cite{meyer_2002, benedetti_1998, capone_2006}. We then have $E_{b,c1}/D e^{-(g/\omega_0)^2} = 2.5$ and $E_{b,c2}/D e^{-(g/\omega_0)^2 }= 3.0$.
These estimates for $g_{c1}$ and $g_{c2}$ may be expressed, with the definition $x=2\omega_0/U_c$ and $y=E_{b,c}/U_c$, where the subscript $c$ refers to $c1$ or $c2$, by the single scaling function, $y/\ln(y) = -x$. The numerical results of (a) are plotted on top of this scaling function in the plot (b). Their reasonable overlaps indicate that the polaronic effects play an important role.

{\it Superconducting $\Delta_p$ and $T_c$ }
--
The outcomes of the DMFT calculations are the diagonal self-energy $\Sigma(\omega)$ and off-diagonal self-energy $\phi(\omega)$ in SC state. There is no renormalization of the chemical potential $\mu$ because at the half-filling it is fixed at the center of the band. We will present below the calculations for $T_c$, $\Delta_p$, density of states, and the spectral function $A(\epsilon,\omega)$ from the obtained self-energies.

The SC gap $\Delta_p$ is given by the solution of
 \ba
 Re \Delta(\omega=\Delta_p) = \Delta_p,
 \ea
where $\Delta(\omega) = \phi(\omega)/Z(\omega)$ and $Z(\omega) = 1-\Sigma(\omega)/\omega $. They are shown in Fig.\ 2.
One can see that for the anti-adiabatic case of $\omega_0 = 2D$ shown in the plot (a), $T_c$ increases and have a broad peak and decreases, but $\Delta_p$ keeps increasing as $g$ increases like the attractive-Hubbard model. On the other hand, for the adiabatic case of $\omega_0 = 0.1 D$ in plot (b), both $T_c$ and $\Delta_p$ increase and rapidly decrease as $g$ increases. This behavior may be understood by noticing that the bandwidth in the Holstein model is scaled down by the factor of $e^{-(g/\omega_0)^2}$ as discussed above.

$T_c$ may be determined by fitting the NRG energy flows ($E_N/\Lambda^{-N/2}$) with respect to the NRG site number  $N$ as $\Lambda^{(N-N_{Tc})/2}$, and the $T_c$ is given by
 \ba
 T_c=\Lambda^{-N_{Tc}/2},
 \ea
where $\Lambda$ is the NRG discretization parameter \cite{hecht_2008}. The NRG-DMFT calculations of $T_c$ are shown with the black asterisks. See the Supplementary Material for more details. The ratio $2\Delta_p/T_c$ is also plotted. It shows the BCS value of 3.5 for small $g$ and continues to increase as $g$ increases.

Recall that $T_c$ is limited by $\Delta_p$ in the BCS and by $D_S$ in BEC regimes, where $D_S$ is the superfluid stiffness. An estimate of $T_c$ taking this observation into account may be done by an interpolation of the gap amplitude and phase stiffness temperature scales of $T_\Delta=\Delta_p/1.75$ and $T_D=D_S/2$ such as $ T_c = (T_{\Delta}^{-\alpha}+T_{D}^{-\alpha})^{-1/\alpha}$. We
found that the exponent $\alpha=1.4$ gives the best results from the least square fitting of the above NRG $T_c$ calculations. This was represented by the gray dashed curves in Fig.\ 2 which reasonably reproduce the numerical calculations.

\begin{figure}
\begin{center}
\includegraphics[scale=0.75]{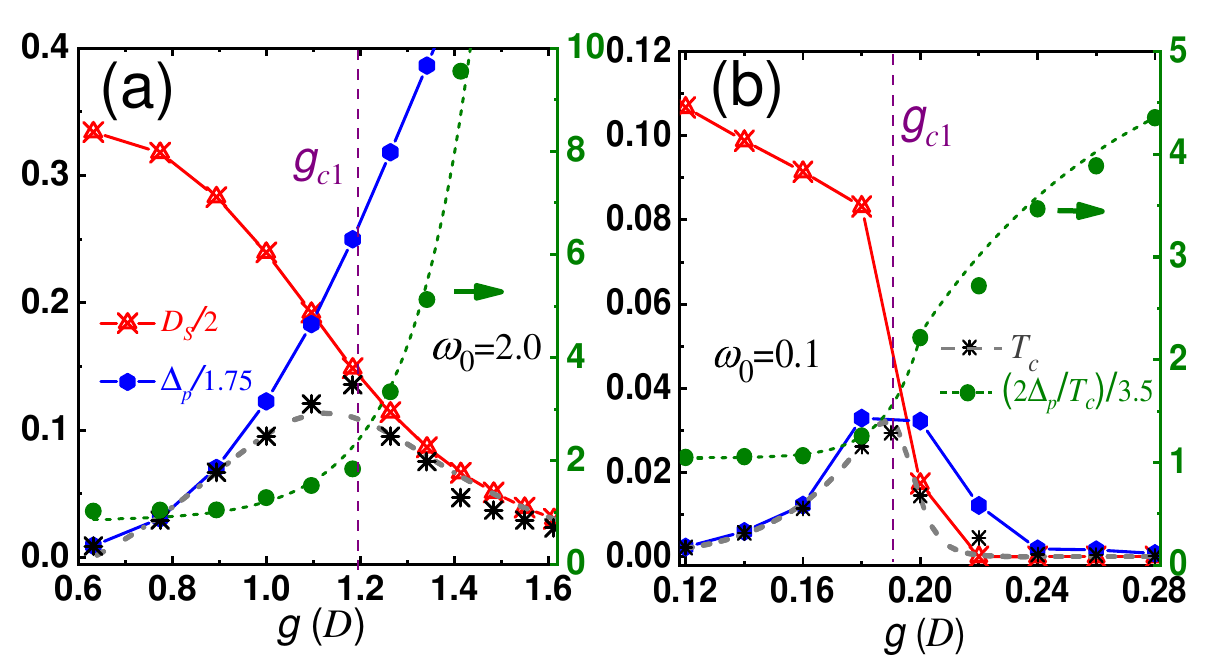}
 \caption{ NRG-DMFT results of the pairing gap $\Delta_p$ (blue circle),
 the superfluid stiffness $D_S$ (red cross-triangle), the critical temperature $T_c$ (black asterisk), and $2\Delta_p/T_c$ (green solid circles)
 for (a) anti-adiabatic and (b) adiabatic cases.
 The black dashed line is the estimate of $T_c$ from $\Delta_p$ and $D_S$.
 Notice that $T_c$ is maximum at $g_{c1}$ for both anti-adiabatic and adiabatic cases.}
\label{fig:gap_stiffness}
\end{center}
\end{figure}

The most interesting for $T_c$ is that the maximum occurs at the lower critical value of $g_{c1}$. This holds regardless of the phonon frequency as can be seen from the Fig.\ 2. It was already noticed in the attractive Hubbard model by Toschi et al.\ that the maximum $T_c$ occurs at $U_{c1}$ \cite{toschi_2005b,kuleeva_2014}.
One may expect that $T_c$ becomes maximum around the crossover regime where the $T_c$ limiting temperature scales $T_\Delta$ and $T_D$ become comparable. The present calculations show that the maximum $T_c$
occurs precisely at $g_{c1}$ where appears the soft phonon mode associated
with the bipolaron instability.

As the electron-phonon coupling $g$ increases the phonon frequency is renormalized and a zero frequency mode appears at the critical coupling. As $g$ is further increased it hardens back in normal state \cite{jeon_2004}.
In SC state, however, for $g$ larger than the critical coupling (BEC regime) the zero frequency weight remains non-vanishing. This is due to the Goldstone mode of the phase fluctuations of local pairs \cite{ranninger_1990}. The last term in the Holstein model of Eq.\ (1) describes the coupling between the phonons and the local electron density. The phonon spectrum therefore reflects the local pairs in the BEC state. Fuller discussions on the Goldstone mode, phonon spectra, and their connection with the maximum $T_c$ will be reported separately.

$Dynamical ~properties$
--
The electronic density of states in SC states are presented in Fig.\ 3. Plot (a) is for the anti-adiabatic and (b) is for the adiabatic cases. If the system is metallic when SC is turned off (BCS regime), then DOS exhibits only one peak (for $\omega>0$) at $\omega=\Delta_p$ in SC state. On the other hand, if insulating when SC is turned off (BEC regime), DOS may exhibit two peaks in SC state: one at $\omega=\Delta_p$ from SC and the other around $\omega \approx E_b/2$ from the insulating gap. These two peaks clearly show up only around crossover regime for the adiabatic case as can be seen from the plot (b) for $g/D=0.2$. For other regimes only single peak shows up in DOS either because the two energy scales merge into one (anti-adiabatic case) or the pairing coherence peak is suppressed due to the small superfluid stiffness (adiabatic case).

\begin{figure}
\begin{center}
\includegraphics[scale=0.6]{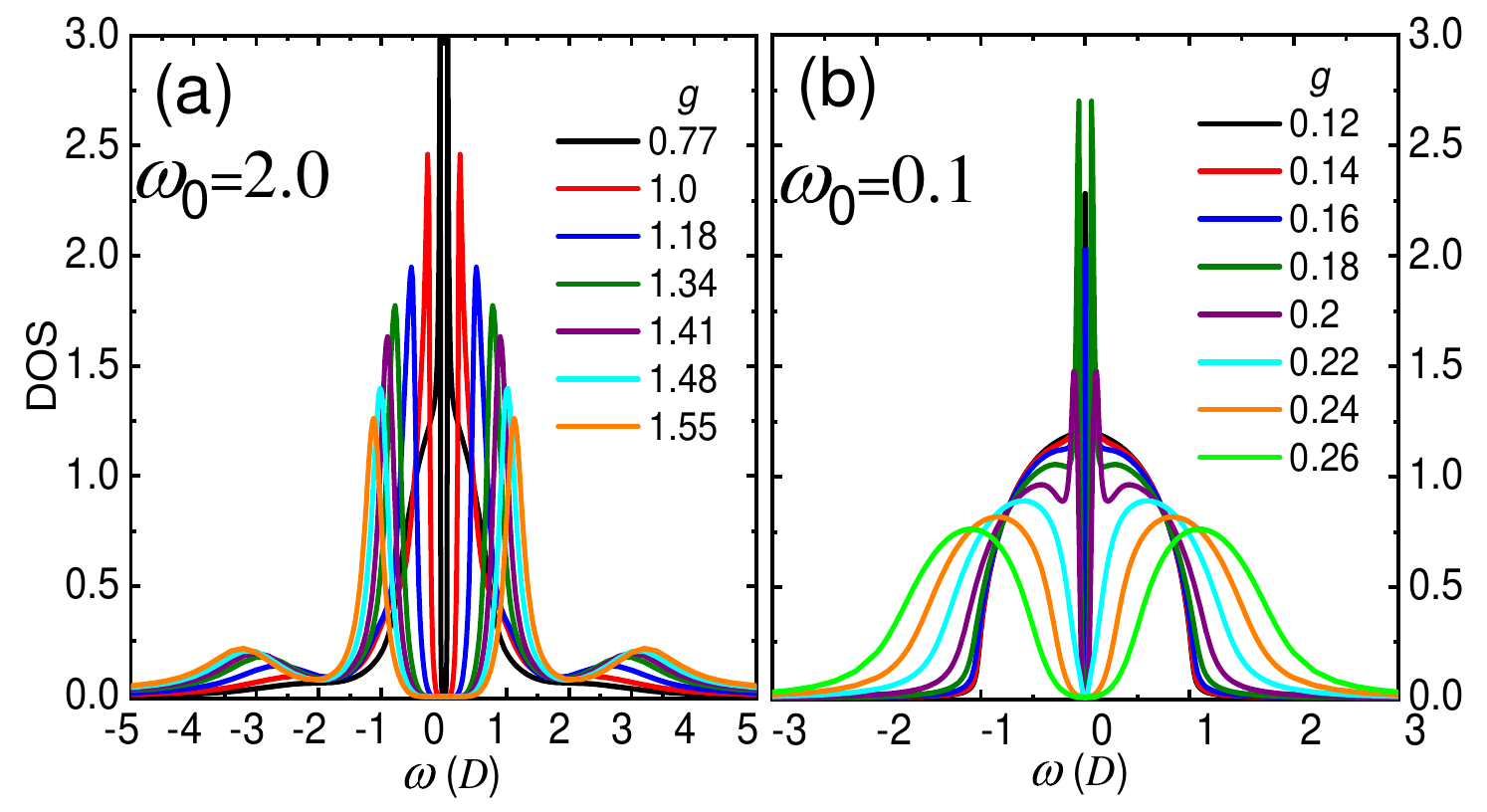}
 \caption{ Electronic density of states of the Holstein model in SC state. (a) is for the anti-adiabatic and (b) is for the adiabatic cases.
 }
\label{fig:eDOS}
\end{center}
\end{figure}

The phase coherence is due to the superfluid stiffness, and we already
saw that $D_S$ is significantly suppressed
in the BEC regime of the adiabatic case from Fig.\ \ref{fig:gap_stiffness}(b).
However, $D_S$ does not decrease rapidly in the anti-adiabatic case as shown
in Fig.\ \ref{fig:gap_stiffness}(a). Therefore, as shown in Fig.\ \ref{fig:eDOS}(a), the coherence peaks remain unsuppressed
together with the local pair states at $ \omega \approx\Delta_p \approx E_b/2$ in BEC regime.
The higher order peaks also appear at the addition of phonon energy $\omega_0$.

\begin{figure}
\begin{center}
\includegraphics[scale=0.6]{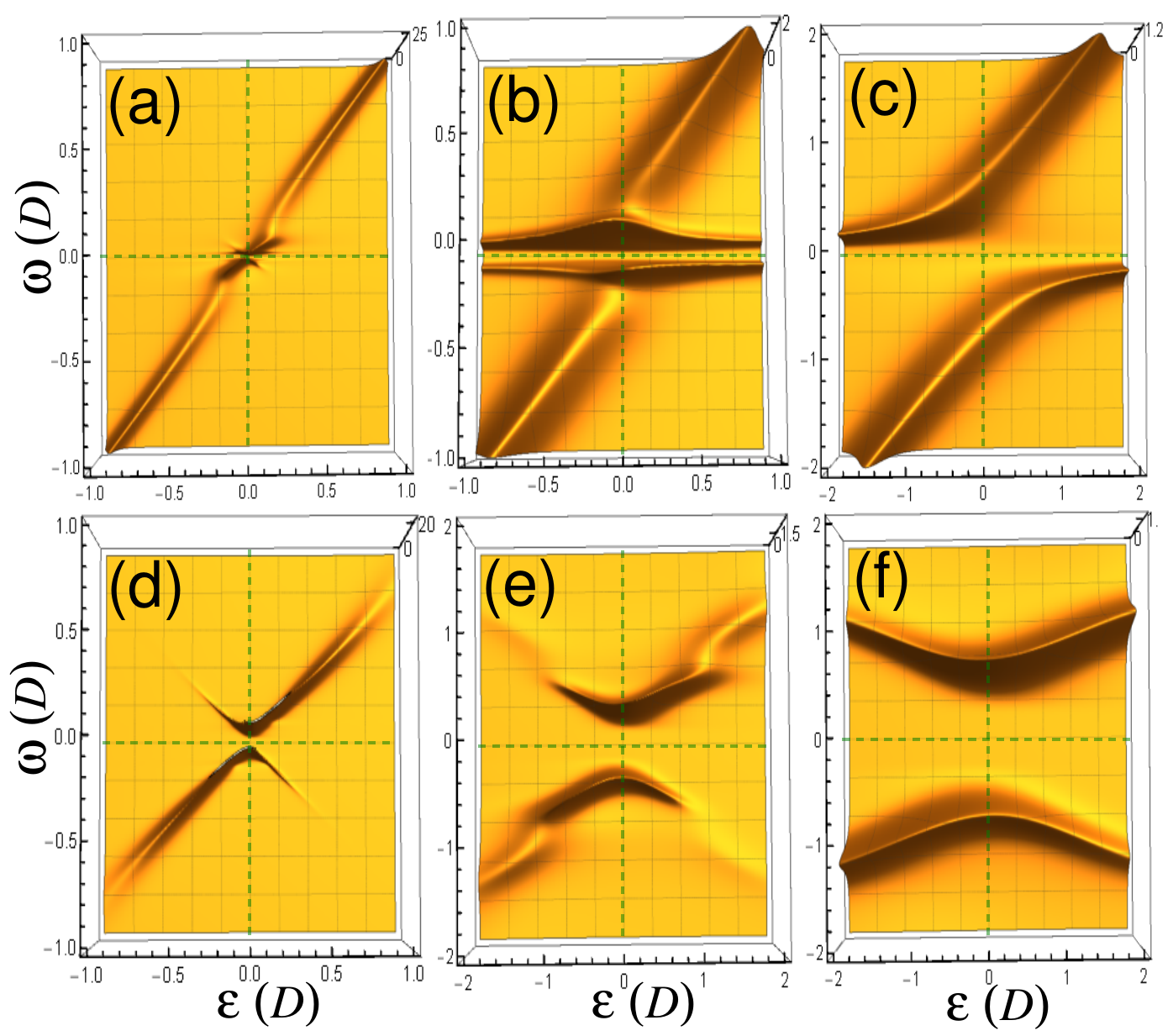}
 \caption{ The spectral intensity as a function of $\epsilon$ and $\omega$ for adiabatic (upper) and anti-adiabatic (lower) cases in the BCS (left), crossover (midle), and BEC (right) regimes. The electron-phonon couplings, $g/D$, is $0.16$, $0.2$, and $0.24$ for (a), (b), and (c), and $g_{c1} \approx 0.19D$. $g/D$ is $0.77$, $1.1$, and $1.41$ for (d), (e), and (f), respectively, with $g_{c1} \approx 1.2D$. Notice the absence of back-bending in the crossover and BEC regimes for adiabatic case in (b) and (c). }
\label{fig:ARPES}
\end{center}
\end{figure}

$Spectral~ intensity~ A(\epsilon,\omega)$
--
We now show the spectral function in Fig.\ 4.
 \ba
A(\epsilon,\omega)=-\frac1\pi \text{Im} G(\epsilon,\omega),
 \ea
where, $\epsilon$ stands for $\epsilon_k$. The spectral intensity plot in the plane of of $\epsilon$
and $\omega$ represents the quasiparticle dispersion relation, which is probed by the angle-resolved
photoemision spectroscopy (ARPES) experiments. We present the representative spectral intensity plots
for BCS, crossover, and BEC regimes, respectively, in Fig.\ \ref{fig:ARPES}(a), (b), and (c) for the adiabatic case, and in (d), (e), and (f) for the anti-adiabatic case. The bare dispersion is along the line of $\omega=\epsilon$.
In the BCS regime, the back-bending
behaviors around the Fermi surface ($\epsilon=0$) can clearly be seen for both adiabatic (plot (a)) and anti-adiabatic (plot (d)) cases. The kink around the bare phonon frequency
($\omega \sim \omega_0$) is also observable for the adiabatic phonon in accord with the mean-field theory calculation of Sandvik {\it et al} \cite{sandvik_2004}.

In the BEC regime, the quasi-particle dispersion for adiabatic and anti-adiabatic cases
behave differently as shown in Fig.\ \ref{fig:ARPES}(c) and (f). For the adiabatic case, the dispersion curve, as $\epsilon$ increases,
passes through the Fermi surface without the back-bending and resumes the normal state dispersion ($\omega=\epsilon$). The presumed back-bending feature (in, for example, $\epsilon>0$ and $\omega<0$ quadrant) is absent which appears in all regimes in anti-adiabatic case. Instead, the dispersion gap keeps decreasing
as one moves away from the bare Fermi surface.
The absence of the back-bending was reported by Rinott {\it et al} by the ARPES experiments
on $\text{Fe}_{1+y}\text{Se}_x\text{Te}_{1-x}$ as mentioned in introduction \cite{rinott_2017}.

$Summary ~and ~discussion$\label{sec:discussion}
--
Our main results for dynamical effects on the crossover are: (a) The pairing gap $\Delta_p$ does not keep increasing; $\Delta_p$ increases and decreases as the coupling $g$ increases for adiabatic phonon cases. (b) The maximum $T_c$ occurs at the lower critical coupling $g_{c1}$ concomitantly with the emergence of the Goldstone mode due to the local pair phase fluctuations. (c) The SC quasi-particle dispersion in the crossover or BEC regime does not exhibit the back-bending in the adiabatic case.
The (b) holds regardless of the phonon frequency, and (a) and (c) are pronounced in the adiabatic cases.

It is important to note that the quasi-particle dispersion may or may not exhibit the back-bending in the BEC regime of superconductivity from dynamically induced insulators. In contrast, the back-bending always shows up in superconductivity emerging from insulators induced by an instantaneous interaction like the attractive-Hubbard model, while it is absent in superconductivity from band insulators \cite{Loh_2016}.
The absence of SC back-bending in latter is understood in the simple mean-field theory picture as follows: Suppose a band insulator where the chemical potential is below the bottom of a conduction band and the minimum gap position is at $k=0$. In SC state, this is in BEC regime and the dispersion gap minimum also occurs at $k=0$. The topology of the  minimum gap loci is a finite volume contour for BCS but a point in ${\bf k}$-space in BEC regimes. However, in the present problem the chemical potential is always at the center of a particle-hole symmetric band to ensure the half-filling, and above picture of the chemical potential out of a band can not be applied.

Superconductivity emerging from a dynamically induced insulator we consider here, the absence of back-bending can be traced back to the insulating behavior of the diagonal self-energy $\Sigma(\omega)$ in SC state, that is, the slope of $Re \Sigma(\omega) \rightarrow \infty$ as $\omega\rightarrow 0$. In both adiabatic and anti-adiabatic cases the self-energy in the local pair insulating regime shows this behavior in the normal state like in the Mott insulating state \cite{bulla_2001}. In SC state, however, the low energy fluctuations leading to insulating behavior are blocked out in anti-adiabatic cases by the large energy gap of $\Delta_p$. But this blocking is not complete in adiabatic cases and the insulating behavior survives. Like the band insulator cases, this insulating behavior underlies the absence of the back-bending. See the Supplementary Material for more details. Application of this idea to Fe$_{1+y}$Se$_x$Te$_{1-x}$ needs more considerations. In particular it seems that the multigap nature of the material ought to be considered in the context of dynamical effects on superconductivity. It will be interesting to observe the BCS-BEC crossover out of interaction induced insulators.

$Acknowledgment$
--
We acknowledge the supports from Samsung Science \& Technology Foundation through SSTF-BA1502-06, and from National Research Foundation of Korea through NRF-2016R1A6A3A11930653 (THP) and NRF-2018R1D1A1B07043997 (HYC).

\bibliography{reference.bib}{}

\begin{thebibliography}{43}%
\makeatletter
\providecommand \@ifxundefined [1]{%
 \@ifx{#1\undefined}
}%
\providecommand \@ifnum [1]{%
 \ifnum #1\expandafter \@firstoftwo
 \else \expandafter \@secondoftwo
 \fi
}%
\providecommand \@ifx [1]{%
 \ifx #1\expandafter \@firstoftwo
 \else \expandafter \@secondoftwo
 \fi
}%
\providecommand \natexlab [1]{#1}%
\providecommand \enquote  [1]{``#1''}%
\providecommand \bibnamefont  [1]{#1}%
\providecommand \bibfnamefont [1]{#1}%
\providecommand \citenamefont [1]{#1}%
\providecommand \href@noop [0]{\@secondoftwo}%
\providecommand \href [0]{\begingroup \@sanitize@url \@href}%
\providecommand \@href[1]{\@@startlink{#1}\@@href}%
\providecommand \@@href[1]{\endgroup#1\@@endlink}%
\providecommand \@sanitize@url [0]{\catcode `\\12\catcode `\$12\catcode
  `\&12\catcode `\#12\catcode `\^12\catcode `\_12\catcode `\%12\relax}%
\providecommand \@@startlink[1]{}%
\providecommand \@@endlink[0]{}%
\providecommand \url  [0]{\begingroup\@sanitize@url \@url }%
\providecommand \@url [1]{\endgroup\@href {#1}{\urlprefix }}%
\providecommand \urlprefix  [0]{URL }%
\providecommand \Eprint [0]{\href }%
\providecommand \doibase [0]{http://dx.doi.org/}%
\providecommand \selectlanguage [0]{\@gobble}%
\providecommand \bibinfo  [0]{\@secondoftwo}%
\providecommand \bibfield  [0]{\@secondoftwo}%
\providecommand \translation [1]{[#1]}%
\providecommand \BibitemOpen [0]{}%
\providecommand \bibitemStop [0]{}%
\providecommand \bibitemNoStop [0]{.\EOS\space}%
\providecommand \EOS [0]{\spacefactor3000\relax}%
\providecommand \BibitemShut  [1]{\csname bibitem#1\endcsname}%
\let\auto@bib@innerbib\@empty
\bibitem [{\citenamefont {Griffin}\ \emph {et~al.}(1994)\citenamefont
  {Griffin}, \citenamefont {Snoke},\ and\ \citenamefont
  {Stringari}}]{griffin_1994}%
  \BibitemOpen
  \bibfield  {author} {\bibinfo {author} {\bibfnamefont {A.}~\bibnamefont
  {Griffin}}, \bibinfo {author} {\bibfnamefont {D.}~\bibnamefont {Snoke}}, \
  and\ \bibinfo {author} {\bibfnamefont {S.}~\bibnamefont {Stringari}},\
  }\href@noop {} {\emph {\bibinfo {title} {Bose-Einstein Condensation}}}\
  (\bibinfo  {publisher} {Cambridge University Press},\ \bibinfo {address}
  {Cambridge, England},\ \bibinfo {year} {1994})\BibitemShut {NoStop}%
\bibitem [{\citenamefont {Leggett}\ and\ \citenamefont
  {Zhang}(2012)}]{leggett_2012}%
  \BibitemOpen
  \bibfield  {author} {\bibinfo {author} {\bibfnamefont {A.~J.}\ \bibnamefont
  {Leggett}}\ and\ \bibinfo {author} {\bibfnamefont {S.}~\bibnamefont
  {Zhang}},\ }\enquote {\bibinfo {title} {The bec--bcs crossover: Some history
  and some general observations},}\ in\ \href {\doibase
  10.1007/978-3-642-21978-8_2} {\emph {\bibinfo {booktitle} {The BCS-BEC
  Crossover and the Unitary Fermi Gas}}},\ \bibinfo {editor} {edited by\
  \bibinfo {editor} {\bibfnamefont {Wilhelm}\ \bibnamefont {Zwerger}}}\
  (\bibinfo  {publisher} {Springer Berlin Heidelberg},\ \bibinfo {address}
  {Berlin, Heidelberg},\ \bibinfo {year} {2012})\ pp.\ \bibinfo {pages}
  {33--47}\BibitemShut {NoStop}%
\bibitem [{\citenamefont {Randeria}\ and\ \citenamefont
  {Taylor}(2014)}]{randeria_2014}%
  \BibitemOpen
  \bibfield  {author} {\bibinfo {author} {\bibfnamefont {Mohit}\ \bibnamefont
  {Randeria}}\ and\ \bibinfo {author} {\bibfnamefont {Edward}\ \bibnamefont
  {Taylor}},\ }\bibfield  {title} {\enquote {\bibinfo {title} {Crossover from
  bardeen-cooper-schrieffer to bose-einstein condensation and the unitary fermi
  gas},}\ }\href {\doibase 10.1146/annurev-conmatphys-031113-133829} {\bibfield
   {journal} {\bibinfo  {journal} {Annual Review of Condensed Matter Physics}\
  }\textbf {\bibinfo {volume} {5}},\ \bibinfo {pages} {209--232} (\bibinfo
  {year} {2014})}\BibitemShut {NoStop}%
\bibitem [{\citenamefont {Eagles}(1969)}]{eagles_1969}%
  \BibitemOpen
  \bibfield  {author} {\bibinfo {author} {\bibfnamefont {D.~M.}\ \bibnamefont
  {Eagles}},\ }\bibfield  {title} {\enquote {\bibinfo {title} {Possible pairing
  without superconductivity at low carrier concentrations in bulk and thin-film
  superconducting semiconductors},}\ }\href {\doibase 10.1103/PhysRev.186.456}
  {\bibfield  {journal} {\bibinfo  {journal} {Phys. Rev.}\ }\textbf {\bibinfo
  {volume} {186}},\ \bibinfo {pages} {456--463} (\bibinfo {year}
  {1969})}\BibitemShut {NoStop}%
\bibitem [{\citenamefont {Leggett}(1980)}]{leggett_1980}%
  \BibitemOpen
  \bibfield  {author} {\bibinfo {author} {\bibfnamefont {A.~J.}\ \bibnamefont
  {Leggett}},\ }\enquote {\bibinfo {title} {Diatomic molecules and cooper
  pairs},}\ in\ \href@noop {} {\emph {\bibinfo {booktitle} {Modern Trends in
  the Theory of Condensed Matter. Proceedings of the XVIth Karpacz Winter
  School of Theoretical Physics, Karpacz, Poland}}}\ (\bibinfo  {publisher}
  {Springer-Verlag},\ \bibinfo {year} {1980})\ pp.\ \bibinfo {pages}
  {13--27}\BibitemShut {NoStop}%
\bibitem [{\citenamefont {Nozi{\`e}res}\ and\ \citenamefont
  {Schmitt-Rink}(1985)}]{nozieres_1985}%
  \BibitemOpen
  \bibfield  {author} {\bibinfo {author} {\bibfnamefont {P.}~\bibnamefont
  {Nozi{\`e}res}}\ and\ \bibinfo {author} {\bibfnamefont {S.}~\bibnamefont
  {Schmitt-Rink}},\ }\bibfield  {title} {\enquote {\bibinfo {title} {Bose
  condensation in an attractive fermion gas: From weak to strong coupling
  superconductivity},}\ }\href {\doibase 10.1007/BF00683774} {\bibfield
  {journal} {\bibinfo  {journal} {Journal of Low Temperature Physics}\ }\textbf
  {\bibinfo {volume} {59}},\ \bibinfo {pages} {195--211} (\bibinfo {year}
  {1985})}\BibitemShut {NoStop}%
\bibitem [{\citenamefont {Randeria}\ \emph {et~al.}(1989)\citenamefont
  {Randeria}, \citenamefont {Duan},\ and\ \citenamefont
  {Shieh}}]{randeria_1989}%
  \BibitemOpen
  \bibfield  {author} {\bibinfo {author} {\bibfnamefont {Mohit}\ \bibnamefont
  {Randeria}}, \bibinfo {author} {\bibfnamefont {Ji-Min}\ \bibnamefont {Duan}},
  \ and\ \bibinfo {author} {\bibfnamefont {Lih-Yir}\ \bibnamefont {Shieh}},\
  }\bibfield  {title} {\enquote {\bibinfo {title} {Bound states, cooper
  pairing, and bose condensation in two dimensions},}\ }\href {\doibase
  10.1103/PhysRevLett.62.981} {\bibfield  {journal} {\bibinfo  {journal} {Phys.
  Rev. Lett.}\ }\textbf {\bibinfo {volume} {62}},\ \bibinfo {pages} {981--984}
  (\bibinfo {year} {1989})}\BibitemShut {NoStop}%
\bibitem [{\citenamefont {S\'a~de Melo}\ \emph {et~al.}(1993)\citenamefont
  {S\'a~de Melo}, \citenamefont {Randeria},\ and\ \citenamefont
  {Engelbrecht}}]{melo_1993}%
  \BibitemOpen
  \bibfield  {author} {\bibinfo {author} {\bibfnamefont {C.~A.~R.}\
  \bibnamefont {S\'a~de Melo}}, \bibinfo {author} {\bibfnamefont {Mohit}\
  \bibnamefont {Randeria}}, \ and\ \bibinfo {author} {\bibfnamefont {Jan~R.}\
  \bibnamefont {Engelbrecht}},\ }\bibfield  {title} {\enquote {\bibinfo {title}
  {Crossover from bcs to bose superconductivity: Transition temperature and
  time-dependent ginzburg-landau theory},}\ }\href {\doibase
  10.1103/PhysRevLett.71.3202} {\bibfield  {journal} {\bibinfo  {journal}
  {Phys. Rev. Lett.}\ }\textbf {\bibinfo {volume} {71}},\ \bibinfo {pages}
  {3202--3205} (\bibinfo {year} {1993})}\BibitemShut {NoStop}%
\bibitem [{\citenamefont {Engelbrecht}\ \emph {et~al.}(1997)\citenamefont
  {Engelbrecht}, \citenamefont {Randeria},\ and\ \citenamefont
  {S\'ade~Melo}}]{engelbrecht_1997}%
  \BibitemOpen
  \bibfield  {author} {\bibinfo {author} {\bibfnamefont {Jan~R.}\ \bibnamefont
  {Engelbrecht}}, \bibinfo {author} {\bibfnamefont {Mohit}\ \bibnamefont
  {Randeria}}, \ and\ \bibinfo {author} {\bibfnamefont {C.~A.~R.}\ \bibnamefont
  {S\'ade~Melo}},\ }\bibfield  {title} {\enquote {\bibinfo {title} {Bcs to bose
  crossover: Broken-symmetry state},}\ }\href {\doibase
  10.1103/PhysRevB.55.15153} {\bibfield  {journal} {\bibinfo  {journal} {Phys.
  Rev. B}\ }\textbf {\bibinfo {volume} {55}},\ \bibinfo {pages} {15153--15156}
  (\bibinfo {year} {1997})}\BibitemShut {NoStop}%
\bibitem [{\citenamefont {Chen}\ \emph {et~al.}(2005)\citenamefont {Chen},
  \citenamefont {Stajic}, \citenamefont {Tan},\ and\ \citenamefont
  {Levin}}]{chen_2005}%
  \BibitemOpen
  \bibfield  {author} {\bibinfo {author} {\bibfnamefont {Qijin}\ \bibnamefont
  {Chen}}, \bibinfo {author} {\bibfnamefont {Jelena}\ \bibnamefont {Stajic}},
  \bibinfo {author} {\bibfnamefont {Shina}\ \bibnamefont {Tan}}, \ and\
  \bibinfo {author} {\bibfnamefont {K.}~\bibnamefont {Levin}},\ }\bibfield
  {title} {\enquote {\bibinfo {title} {Bcs–bec crossover: From high
  temperature superconductors to ultracold superfluids},}\ }\href {\doibase
  https://doi.org/10.1016/j.physrep.2005.02.005} {\bibfield  {journal}
  {\bibinfo  {journal} {Physics Reports}\ }\textbf {\bibinfo {volume} {412}},\
  \bibinfo {pages} {1 -- 88} (\bibinfo {year} {2005})}\BibitemShut {NoStop}%
\bibitem [{\citenamefont {Lee}\ \emph {et~al.}(2006)\citenamefont {Lee},
  \citenamefont {Nagaosa},\ and\ \citenamefont {Wen}}]{Lee_2006}%
  \BibitemOpen
  \bibfield  {author} {\bibinfo {author} {\bibfnamefont {Patrick~A.}\
  \bibnamefont {Lee}}, \bibinfo {author} {\bibfnamefont {Naoto}\ \bibnamefont
  {Nagaosa}}, \ and\ \bibinfo {author} {\bibfnamefont {Xiao-Gang}\ \bibnamefont
  {Wen}},\ }\bibfield  {title} {\enquote {\bibinfo {title} {Doping a mott
  insulator: Physics of high-temperature superconductivity},}\ }\href {\doibase
  10.1103/RevModPhys.78.17} {\bibfield  {journal} {\bibinfo  {journal} {Rev.
  Mod. Phys.}\ }\textbf {\bibinfo {volume} {78}},\ \bibinfo {pages} {17--85}
  (\bibinfo {year} {2006})}\BibitemShut {NoStop}%
\bibitem [{\citenamefont {Gaebler}\ \emph {et~al.}(2010)\citenamefont
  {Gaebler}, \citenamefont {Stewart}, \citenamefont {Drake}, \citenamefont
  {Jin}, \citenamefont {Perali}, \citenamefont {Pieri},\ and\ \citenamefont
  {Strinati}}]{gaebler_2010}%
  \BibitemOpen
  \bibfield  {author} {\bibinfo {author} {\bibfnamefont {J~P}\ \bibnamefont
  {Gaebler}}, \bibinfo {author} {\bibfnamefont {J~T}\ \bibnamefont {Stewart}},
  \bibinfo {author} {\bibfnamefont {T~E}\ \bibnamefont {Drake}}, \bibinfo
  {author} {\bibfnamefont {D~S}\ \bibnamefont {Jin}}, \bibinfo {author}
  {\bibfnamefont {A}~\bibnamefont {Perali}}, \bibinfo {author} {\bibfnamefont
  {P}~\bibnamefont {Pieri}}, \ and\ \bibinfo {author} {\bibfnamefont {G~C}\
  \bibnamefont {Strinati}},\ }\bibfield  {title} {\enquote {\bibinfo {title}
  {{Observation of pseudogap behaviour in a strongly interacting Fermi gas}},}\
  }\href {http://dx.doi.org/10.1038/nphys1709 http://10.0.4.14/nphys1709}
  {\bibfield  {journal} {\bibinfo  {journal} {Nature Physics}\ }\textbf
  {\bibinfo {volume} {6}},\ \bibinfo {pages} {569} (\bibinfo {year}
  {2010})}\BibitemShut {NoStop}%
\bibitem [{\citenamefont {Regal}\ \emph {et~al.}(2005)\citenamefont {Regal},
  \citenamefont {Greiner}, \citenamefont {Giorgini}, \citenamefont {Holland},\
  and\ \citenamefont {Jin}}]{regal_2005}%
  \BibitemOpen
  \bibfield  {author} {\bibinfo {author} {\bibfnamefont {C.~A.}\ \bibnamefont
  {Regal}}, \bibinfo {author} {\bibfnamefont {M.}~\bibnamefont {Greiner}},
  \bibinfo {author} {\bibfnamefont {S.}~\bibnamefont {Giorgini}}, \bibinfo
  {author} {\bibfnamefont {M.}~\bibnamefont {Holland}}, \ and\ \bibinfo
  {author} {\bibfnamefont {D.~S.}\ \bibnamefont {Jin}},\ }\bibfield  {title}
  {\enquote {\bibinfo {title} {Momentum distribution of a fermi gas of atoms in
  the bcs-bec crossover},}\ }\href {\doibase 10.1103/PhysRevLett.95.250404}
  {\bibfield  {journal} {\bibinfo  {journal} {Phys. Rev. Lett.}\ }\textbf
  {\bibinfo {volume} {95}},\ \bibinfo {pages} {250404} (\bibinfo {year}
  {2005})}\BibitemShut {NoStop}%
\bibitem [{\citenamefont {Bloch}\ \emph {et~al.}(2008)\citenamefont {Bloch},
  \citenamefont {Dalibard},\ and\ \citenamefont {Zwerger}}]{bloch_2008}%
  \BibitemOpen
  \bibfield  {author} {\bibinfo {author} {\bibfnamefont {Immanuel}\
  \bibnamefont {Bloch}}, \bibinfo {author} {\bibfnamefont {Jean}\ \bibnamefont
  {Dalibard}}, \ and\ \bibinfo {author} {\bibfnamefont {Wilhelm}\ \bibnamefont
  {Zwerger}},\ }\bibfield  {title} {\enquote {\bibinfo {title} {Many-body
  physics with ultracold gases},}\ }\href {\doibase 10.1103/RevModPhys.80.885}
  {\bibfield  {journal} {\bibinfo  {journal} {Rev. Mod. Phys.}\ }\textbf
  {\bibinfo {volume} {80}},\ \bibinfo {pages} {885--964} (\bibinfo {year}
  {2008})}\BibitemShut {NoStop}%
\bibitem [{\citenamefont {Chin}\ \emph {et~al.}(2010)\citenamefont {Chin},
  \citenamefont {Grimm}, \citenamefont {Julienne},\ and\ \citenamefont
  {Tiesinga}}]{chin_2010}%
  \BibitemOpen
  \bibfield  {author} {\bibinfo {author} {\bibfnamefont {Cheng}\ \bibnamefont
  {Chin}}, \bibinfo {author} {\bibfnamefont {Rudolf}\ \bibnamefont {Grimm}},
  \bibinfo {author} {\bibfnamefont {Paul}\ \bibnamefont {Julienne}}, \ and\
  \bibinfo {author} {\bibfnamefont {Eite}\ \bibnamefont {Tiesinga}},\
  }\bibfield  {title} {\enquote {\bibinfo {title} {Feshbach resonances in
  ultracold gases},}\ }\href {\doibase 10.1103/RevModPhys.82.1225} {\bibfield
  {journal} {\bibinfo  {journal} {Rev. Mod. Phys.}\ }\textbf {\bibinfo {volume}
  {82}},\ \bibinfo {pages} {1225--1286} (\bibinfo {year} {2010})}\BibitemShut
  {NoStop}%
\bibitem [{\citenamefont {Sato}\ \emph {et~al.}(2018)\citenamefont {Sato},
  \citenamefont {Kasahara}, \citenamefont {Taniguchi}, \citenamefont {Xing},
  \citenamefont {Kasahara}, \citenamefont {Tokiwa}, \citenamefont {Yamakawa},
  \citenamefont {Kontani}, \citenamefont {Shibauchi},\ and\ \citenamefont
  {Matsuda}}]{sato_2018}%
  \BibitemOpen
  \bibfield  {author} {\bibinfo {author} {\bibfnamefont {Yuki}\ \bibnamefont
  {Sato}}, \bibinfo {author} {\bibfnamefont {Shigeru}\ \bibnamefont
  {Kasahara}}, \bibinfo {author} {\bibfnamefont {Tomoya}\ \bibnamefont
  {Taniguchi}}, \bibinfo {author} {\bibfnamefont {Xiangzhuo}\ \bibnamefont
  {Xing}}, \bibinfo {author} {\bibfnamefont {Yuichi}\ \bibnamefont {Kasahara}},
  \bibinfo {author} {\bibfnamefont {Yoshifumi}\ \bibnamefont {Tokiwa}},
  \bibinfo {author} {\bibfnamefont {Youichi}\ \bibnamefont {Yamakawa}},
  \bibinfo {author} {\bibfnamefont {Hiroshi}\ \bibnamefont {Kontani}}, \bibinfo
  {author} {\bibfnamefont {Takasada}\ \bibnamefont {Shibauchi}}, \ and\
  \bibinfo {author} {\bibfnamefont {Yuji}\ \bibnamefont {Matsuda}},\ }\bibfield
   {title} {\enquote {\bibinfo {title} {Abrupt change of the superconducting
  gap structure at the nematic critical point in fese1-xsx},}\ }\href {\doibase
  10.1073/pnas.1717331115} {\bibfield  {journal} {\bibinfo  {journal}
  {Proceedings of the National Academy of Sciences}\ }\textbf {\bibinfo
  {volume} {115}},\ \bibinfo {pages} {1227--1231} (\bibinfo {year}
  {2018})}\BibitemShut {NoStop}%
\bibitem [{\citenamefont {Mizukami}\ \emph {et~al.}(2018)\citenamefont
  {Mizukami} \emph {et~al.}}]{mizukami_2018}%
  \BibitemOpen
  \bibfield  {author} {\bibinfo {author} {\bibfnamefont {Yuta}\ \bibnamefont
  {Mizukami}} \emph {et~al.},\ }\bibfield  {title} {\enquote {\bibinfo {title}
  {Abrupt change of the specific heat in fese1-xsx},}\ }\href@noop {}
  {\bibfield  {journal} {\bibinfo  {journal} {unpublished}\ } (\bibinfo {year}
  {2018})}\BibitemShut {NoStop}%
\bibitem [{\citenamefont {Hanaguri}\ \emph {et~al.}(2018)\citenamefont
  {Hanaguri}, \citenamefont {Iwaya}, \citenamefont {Kohsaka}, \citenamefont
  {Machida}, \citenamefont {Watashige}, \citenamefont {Kasahara}, \citenamefont
  {Shibauchi},\ and\ \citenamefont {Matsuda}}]{hanaguri_2018}%
  \BibitemOpen
  \bibfield  {author} {\bibinfo {author} {\bibfnamefont {Tetsuo}\ \bibnamefont
  {Hanaguri}}, \bibinfo {author} {\bibfnamefont {Katsuya}\ \bibnamefont
  {Iwaya}}, \bibinfo {author} {\bibfnamefont {Yuhki}\ \bibnamefont {Kohsaka}},
  \bibinfo {author} {\bibfnamefont {Tadashi}\ \bibnamefont {Machida}}, \bibinfo
  {author} {\bibfnamefont {Tatsuya}\ \bibnamefont {Watashige}}, \bibinfo
  {author} {\bibfnamefont {Shigeru}\ \bibnamefont {Kasahara}}, \bibinfo
  {author} {\bibfnamefont {Takasada}\ \bibnamefont {Shibauchi}}, \ and\
  \bibinfo {author} {\bibfnamefont {Yuji}\ \bibnamefont {Matsuda}},\ }\bibfield
   {title} {\enquote {\bibinfo {title} {Two distinct superconducting pairing
  states divided by the nematic end point in fese1-xsx},}\ }\href {\doibase
  10.1126/sciadv.aar6419} {\bibfield  {journal} {\bibinfo  {journal} {Science
  Advances}\ }\textbf {\bibinfo {volume} {4}} (\bibinfo {year} {2018}),\
  10.1126/sciadv.aar6419}\BibitemShut {NoStop}%
\bibitem [{\citenamefont {Rinott}\ \emph {et~al.}(2017)\citenamefont {Rinott},
  \citenamefont {Chashka}, \citenamefont {Ribak}, \citenamefont {Rienks},
  \citenamefont {Taleb-Ibrahimi}, \citenamefont {Le~Fevre}, \citenamefont
  {Bertran}, \citenamefont {Randeria},\ and\ \citenamefont
  {Kanigel}}]{rinott_2017}%
  \BibitemOpen
  \bibfield  {author} {\bibinfo {author} {\bibfnamefont {Shahar}\ \bibnamefont
  {Rinott}}, \bibinfo {author} {\bibfnamefont {K.~B.}\ \bibnamefont {Chashka}},
  \bibinfo {author} {\bibfnamefont {Amit}\ \bibnamefont {Ribak}}, \bibinfo
  {author} {\bibfnamefont {Emile D.~L.}\ \bibnamefont {Rienks}}, \bibinfo
  {author} {\bibfnamefont {Amina}\ \bibnamefont {Taleb-Ibrahimi}}, \bibinfo
  {author} {\bibfnamefont {Patrick}\ \bibnamefont {Le~Fevre}}, \bibinfo
  {author} {\bibfnamefont {Fran{\c c}ois}\ \bibnamefont {Bertran}}, \bibinfo
  {author} {\bibfnamefont {Mohit}\ \bibnamefont {Randeria}}, \ and\ \bibinfo
  {author} {\bibfnamefont {Amit}\ \bibnamefont {Kanigel}},\ }\bibfield  {title}
  {\enquote {\bibinfo {title} {Tuning across the bcs-bec crossover in the
  multiband superconductor fe1+ysexte1-x: An angle-resolved photoemission
  study},}\ }\href {\doibase 10.1126/sciadv.1602372} {\bibfield  {journal}
  {\bibinfo  {journal} {Science Advances}\ }\textbf {\bibinfo {volume} {3}}
  (\bibinfo {year} {2017}),\ 10.1126/sciadv.1602372}\BibitemShut {NoStop}%
\bibitem [{\citenamefont {Lubashevsky}\ \emph {et~al.}(2012)\citenamefont
  {Lubashevsky}, \citenamefont {Lahoud}, \citenamefont {Chashka}, \citenamefont
  {Podolsky},\ and\ \citenamefont {Kanigel}}]{lubashevsky_2012}%
  \BibitemOpen
  \bibfield  {author} {\bibinfo {author} {\bibfnamefont {Y.}~\bibnamefont
  {Lubashevsky}}, \bibinfo {author} {\bibfnamefont {E.}~\bibnamefont {Lahoud}},
  \bibinfo {author} {\bibfnamefont {K.}~\bibnamefont {Chashka}}, \bibinfo
  {author} {\bibfnamefont {D.}~\bibnamefont {Podolsky}}, \ and\ \bibinfo
  {author} {\bibfnamefont {A.}~\bibnamefont {Kanigel}},\ }\bibfield  {title}
  {\enquote {\bibinfo {title} {Shallow pockets and very strong coupling
  superconductivity in fesexte1-x},}\ }\href {\doibase 10.1038/nphys2216}
  {\bibfield  {journal} {\bibinfo  {journal} {Nature Physics}\ }\textbf
  {\bibinfo {volume} {8}},\ \bibinfo {pages} {309} (\bibinfo {year}
  {2012})}\BibitemShut {NoStop}%
\bibitem [{\citenamefont {Loh}\ \emph {et~al.}(2016)\citenamefont {Loh},
  \citenamefont {Randeria}, \citenamefont {Trivedi}, \citenamefont {Chang},\
  and\ \citenamefont {Scalettar}}]{Loh_2016}%
  \BibitemOpen
  \bibfield  {author} {\bibinfo {author} {\bibfnamefont {Yen~Lee}\ \bibnamefont
  {Loh}}, \bibinfo {author} {\bibfnamefont {Mohit}\ \bibnamefont {Randeria}},
  \bibinfo {author} {\bibfnamefont {Nandini}\ \bibnamefont {Trivedi}}, \bibinfo
  {author} {\bibfnamefont {Chia-Chen}\ \bibnamefont {Chang}}, \ and\ \bibinfo
  {author} {\bibfnamefont {Richard}\ \bibnamefont {Scalettar}},\ }\bibfield
  {title} {\enquote {\bibinfo {title} {Superconductor-insulator transition and
  fermi-bose crossovers},}\ }\href {\doibase 10.1103/PhysRevX.6.021029}
  {\bibfield  {journal} {\bibinfo  {journal} {Phys. Rev. X}\ }\textbf {\bibinfo
  {volume} {6}},\ \bibinfo {pages} {021029} (\bibinfo {year}
  {2016})}\BibitemShut {NoStop}%
\bibitem [{\citenamefont {Benedetti}\ and\ \citenamefont
  {Zeyher}(1998)}]{benedetti_1998}%
  \BibitemOpen
  \bibfield  {author} {\bibinfo {author} {\bibfnamefont {Patrizia}\
  \bibnamefont {Benedetti}}\ and\ \bibinfo {author} {\bibfnamefont {Roland}\
  \bibnamefont {Zeyher}},\ }\bibfield  {title} {\enquote {\bibinfo {title}
  {Holstein model in infinite dimensions at half-filling},}\ }\href {\doibase
  10.1103/PhysRevB.58.14320} {\bibfield  {journal} {\bibinfo  {journal} {Phys.
  Rev. B}\ }\textbf {\bibinfo {volume} {58}},\ \bibinfo {pages} {14320--14334}
  (\bibinfo {year} {1998})}\BibitemShut {NoStop}%
\bibitem [{\citenamefont {Meyer}\ \emph {et~al.}(2002)\citenamefont {Meyer},
  \citenamefont {Hewson},\ and\ \citenamefont {Bulla}}]{meyer_2002}%
  \BibitemOpen
  \bibfield  {author} {\bibinfo {author} {\bibfnamefont {D.}~\bibnamefont
  {Meyer}}, \bibinfo {author} {\bibfnamefont {A.~C.}\ \bibnamefont {Hewson}}, \
  and\ \bibinfo {author} {\bibfnamefont {R.}~\bibnamefont {Bulla}},\ }\bibfield
   {title} {\enquote {\bibinfo {title} {Gap formation and soft phonon mode in
  the holstein model},}\ }\href {\doibase 10.1103/PhysRevLett.89.196401}
  {\bibfield  {journal} {\bibinfo  {journal} {Phys. Rev. Lett.}\ }\textbf
  {\bibinfo {volume} {89}},\ \bibinfo {pages} {196401} (\bibinfo {year}
  {2002})}\BibitemShut {NoStop}%
\bibitem [{\citenamefont {Capone}\ and\ \citenamefont
  {Ciuchi}(2003)}]{capone_2003}%
  \BibitemOpen
  \bibfield  {author} {\bibinfo {author} {\bibfnamefont {M.}~\bibnamefont
  {Capone}}\ and\ \bibinfo {author} {\bibfnamefont {S.}~\bibnamefont
  {Ciuchi}},\ }\bibfield  {title} {\enquote {\bibinfo {title} {Polaron
  crossover and bipolaronic metal-insulator transition in the half-filled
  holstein model},}\ }\href {\doibase 10.1103/PhysRevLett.91.186405} {\bibfield
   {journal} {\bibinfo  {journal} {Phys. Rev. Lett.}\ }\textbf {\bibinfo
  {volume} {91}},\ \bibinfo {pages} {186405} (\bibinfo {year}
  {2003})}\BibitemShut {NoStop}%
\bibitem [{\citenamefont {Jeon}\ \emph {et~al.}(2004)\citenamefont {Jeon},
  \citenamefont {Park}, \citenamefont {Han}, \citenamefont {Lee},\ and\
  \citenamefont {Choi}}]{jeon_2004}%
  \BibitemOpen
  \bibfield  {author} {\bibinfo {author} {\bibfnamefont {Gun~Sang}\
  \bibnamefont {Jeon}}, \bibinfo {author} {\bibfnamefont {Tae-Ho}\ \bibnamefont
  {Park}}, \bibinfo {author} {\bibfnamefont {Jung~Hoon}\ \bibnamefont {Han}},
  \bibinfo {author} {\bibfnamefont {Hyun~C.}\ \bibnamefont {Lee}}, \ and\
  \bibinfo {author} {\bibfnamefont {Han-Yong}\ \bibnamefont {Choi}},\
  }\bibfield  {title} {\enquote {\bibinfo {title} {Dynamical mean-field theory
  of the hubbard-holstein model at half filling: Zero temperature
  metal-insulator and insulator-insulator transitions},}\ }\href {\doibase
  10.1103/PhysRevB.70.125114} {\bibfield  {journal} {\bibinfo  {journal} {Phys.
  Rev. B}\ }\textbf {\bibinfo {volume} {70}},\ \bibinfo {pages} {125114}
  (\bibinfo {year} {2004})}\BibitemShut {NoStop}%
\bibitem [{\citenamefont {Koller}\ \emph {et~al.}(2004)\citenamefont {Koller},
  \citenamefont {Meyer},\ and\ \citenamefont {Hewson}}]{koller_2004}%
  \BibitemOpen
  \bibfield  {author} {\bibinfo {author} {\bibfnamefont {W.}~\bibnamefont
  {Koller}}, \bibinfo {author} {\bibfnamefont {D.}~\bibnamefont {Meyer}}, \
  and\ \bibinfo {author} {\bibfnamefont {A.~C.}\ \bibnamefont {Hewson}},\
  }\bibfield  {title} {\enquote {\bibinfo {title} {Dynamic response functions
  for the holstein-hubbard model},}\ }\href {\doibase
  10.1103/PhysRevB.70.155103} {\bibfield  {journal} {\bibinfo  {journal} {Phys.
  Rev. B}\ }\textbf {\bibinfo {volume} {70}},\ \bibinfo {pages} {155103}
  (\bibinfo {year} {2004})}\BibitemShut {NoStop}%
\bibitem [{\citenamefont {Deppeler}\ and\ \citenamefont
  {Millis}(2002)}]{deppeler_2002}%
  \BibitemOpen
  \bibfield  {author} {\bibinfo {author} {\bibfnamefont {Andreas}\ \bibnamefont
  {Deppeler}}\ and\ \bibinfo {author} {\bibfnamefont {A.~J.}\ \bibnamefont
  {Millis}},\ }\bibfield  {title} {\enquote {\bibinfo {title} {Electron-phonon
  interactions in correlated systems: Adiabatic expansion of the dynamical
  mean-field theory},}\ }\href {\doibase 10.1103/PhysRevB.65.100301} {\bibfield
   {journal} {\bibinfo  {journal} {Phys. Rev. B}\ }\textbf {\bibinfo {volume}
  {65}},\ \bibinfo {pages} {100301} (\bibinfo {year} {2002})}\BibitemShut
  {NoStop}%
\bibitem [{\citenamefont {Murakami}\ \emph {et~al.}(2013)\citenamefont
  {Murakami}, \citenamefont {Werner}, \citenamefont {Tsuji},\ and\
  \citenamefont {Aoki}}]{murakami_2013}%
  \BibitemOpen
  \bibfield  {author} {\bibinfo {author} {\bibfnamefont {Yuta}\ \bibnamefont
  {Murakami}}, \bibinfo {author} {\bibfnamefont {Philipp}\ \bibnamefont
  {Werner}}, \bibinfo {author} {\bibfnamefont {Naoto}\ \bibnamefont {Tsuji}}, \
  and\ \bibinfo {author} {\bibfnamefont {Hideo}\ \bibnamefont {Aoki}},\
  }\bibfield  {title} {\enquote {\bibinfo {title} {Ordered phases in the
  holstein-hubbard model: Interplay of strong coulomb interaction and
  electron-phonon coupling},}\ }\href {\doibase 10.1103/PhysRevB.88.125126}
  {\bibfield  {journal} {\bibinfo  {journal} {Phys. Rev. B}\ }\textbf {\bibinfo
  {volume} {88}},\ \bibinfo {pages} {125126} (\bibinfo {year}
  {2013})}\BibitemShut {NoStop}%
\bibitem [{\citenamefont {Murakami}\ \emph {et~al.}(2014)\citenamefont
  {Murakami}, \citenamefont {Werner}, \citenamefont {Tsuji},\ and\
  \citenamefont {Aoki}}]{murakami_2014}%
  \BibitemOpen
  \bibfield  {author} {\bibinfo {author} {\bibfnamefont {Yuta}\ \bibnamefont
  {Murakami}}, \bibinfo {author} {\bibfnamefont {Philipp}\ \bibnamefont
  {Werner}}, \bibinfo {author} {\bibfnamefont {Naoto}\ \bibnamefont {Tsuji}}, \
  and\ \bibinfo {author} {\bibfnamefont {Hideo}\ \bibnamefont {Aoki}},\
  }\bibfield  {title} {\enquote {\bibinfo {title} {Supersolid phase accompanied
  by a quantum critical point in the intermediate coupling regime of the
  holstein model},}\ }\href {\doibase 10.1103/PhysRevLett.113.266404}
  {\bibfield  {journal} {\bibinfo  {journal} {Phys. Rev. Lett.}\ }\textbf
  {\bibinfo {volume} {113}},\ \bibinfo {pages} {266404} (\bibinfo {year}
  {2014})}\BibitemShut {NoStop}%
\bibitem [{\citenamefont {Bauer}\ \emph {et~al.}(2009)\citenamefont {Bauer},
  \citenamefont {Hewson},\ and\ \citenamefont {Dupuis}}]{bauer_2009}%
  \BibitemOpen
  \bibfield  {author} {\bibinfo {author} {\bibfnamefont {J.}~\bibnamefont
  {Bauer}}, \bibinfo {author} {\bibfnamefont {A.~C.}\ \bibnamefont {Hewson}}, \
  and\ \bibinfo {author} {\bibfnamefont {N.}~\bibnamefont {Dupuis}},\
  }\bibfield  {title} {\enquote {\bibinfo {title} {Dynamical mean-field theory
  and numerical renormalization group study of superconductivity in the
  attractive hubbard model},}\ }\href {\doibase 10.1103/PhysRevB.79.214518}
  {\bibfield  {journal} {\bibinfo  {journal} {Phys. Rev. B}\ }\textbf {\bibinfo
  {volume} {79}},\ \bibinfo {pages} {214518} (\bibinfo {year}
  {2009})}\BibitemShut {NoStop}%
\bibitem [{\citenamefont {Bulla}\ \emph {et~al.}(2008)\citenamefont {Bulla},
  \citenamefont {Costi},\ and\ \citenamefont {Pruschke}}]{bulla_2008}%
  \BibitemOpen
  \bibfield  {author} {\bibinfo {author} {\bibfnamefont {Ralf}\ \bibnamefont
  {Bulla}}, \bibinfo {author} {\bibfnamefont {Theo~A.}\ \bibnamefont {Costi}},
  \ and\ \bibinfo {author} {\bibfnamefont {Thomas}\ \bibnamefont {Pruschke}},\
  }\bibfield  {title} {\enquote {\bibinfo {title} {Numerical renormalization
  group method for quantum impurity systems},}\ }\href {\doibase
  10.1103/RevModPhys.80.395} {\bibfield  {journal} {\bibinfo  {journal} {Rev.
  Mod. Phys.}\ }\textbf {\bibinfo {volume} {80}},\ \bibinfo {pages} {395--450}
  (\bibinfo {year} {2008})}\BibitemShut {NoStop}%
\bibitem [{\citenamefont {Jeon}\ \emph {et~al.}(2003)\citenamefont {Jeon},
  \citenamefont {Park},\ and\ \citenamefont {Choi}}]{jeon_2003}%
  \BibitemOpen
  \bibfield  {author} {\bibinfo {author} {\bibfnamefont {Gun~Sang}\
  \bibnamefont {Jeon}}, \bibinfo {author} {\bibfnamefont {Tae-Ho}\ \bibnamefont
  {Park}}, \ and\ \bibinfo {author} {\bibfnamefont {Han-Yong}\ \bibnamefont
  {Choi}},\ }\bibfield  {title} {\enquote {\bibinfo {title} {Numerical
  renormalization-group study of the symmetric anderson-holstein model: Phonon
  and electron spectral functions},}\ }\href {\doibase
  10.1103/PhysRevB.68.045106} {\bibfield  {journal} {\bibinfo  {journal} {Phys.
  Rev. B}\ }\textbf {\bibinfo {volume} {68}},\ \bibinfo {pages} {045106}
  (\bibinfo {year} {2003})}\BibitemShut {NoStop}%
\bibitem [{\citenamefont {Hewson}\ and\ \citenamefont
  {Meyer}(2002)}]{hewson_2002}%
  \BibitemOpen
  \bibfield  {author} {\bibinfo {author} {\bibfnamefont {A~C}\ \bibnamefont
  {Hewson}}\ and\ \bibinfo {author} {\bibfnamefont {D}~\bibnamefont {Meyer}},\
  }\bibfield  {title} {\enquote {\bibinfo {title} {Numerical renormalization
  group study of the anderson-holstein impurity model},}\ }\href
  {http://stacks.iop.org/0953-8984/14/i=3/a=312} {\bibfield  {journal}
  {\bibinfo  {journal} {Journal of Physics: Condensed Matter}\ }\textbf
  {\bibinfo {volume} {14}},\ \bibinfo {pages} {427} (\bibinfo {year}
  {2002})}\BibitemShut {NoStop}%
\bibitem [{\citenamefont {Paci}\ \emph {et~al.}(2005)\citenamefont {Paci},
  \citenamefont {Capone}, \citenamefont {Cappelluti}, \citenamefont {Ciuchi},
  \citenamefont {Grimaldi},\ and\ \citenamefont {Pietronero}}]{paci_2005}%
  \BibitemOpen
  \bibfield  {author} {\bibinfo {author} {\bibfnamefont {P.}~\bibnamefont
  {Paci}}, \bibinfo {author} {\bibfnamefont {M.}~\bibnamefont {Capone}},
  \bibinfo {author} {\bibfnamefont {E.}~\bibnamefont {Cappelluti}}, \bibinfo
  {author} {\bibfnamefont {S.}~\bibnamefont {Ciuchi}}, \bibinfo {author}
  {\bibfnamefont {C.}~\bibnamefont {Grimaldi}}, \ and\ \bibinfo {author}
  {\bibfnamefont {L.}~\bibnamefont {Pietronero}},\ }\bibfield  {title}
  {\enquote {\bibinfo {title} {Polaronic and nonadiabatic phase diagram from
  anomalous isotope effects},}\ }\href {\doibase 10.1103/PhysRevLett.94.036406}
  {\bibfield  {journal} {\bibinfo  {journal} {Phys. Rev. Lett.}\ }\textbf
  {\bibinfo {volume} {94}},\ \bibinfo {pages} {036406} (\bibinfo {year}
  {2005})}\BibitemShut {NoStop}%
\bibitem [{\citenamefont {Capone}\ \emph {et~al.}(2006)\citenamefont {Capone},
  \citenamefont {Carta},\ and\ \citenamefont {Ciuchi}}]{capone_2006}%
  \BibitemOpen
  \bibfield  {author} {\bibinfo {author} {\bibfnamefont {M.}~\bibnamefont
  {Capone}}, \bibinfo {author} {\bibfnamefont {P.}~\bibnamefont {Carta}}, \
  and\ \bibinfo {author} {\bibfnamefont {S.}~\bibnamefont {Ciuchi}},\
  }\bibfield  {title} {\enquote {\bibinfo {title} {Dynamical mean field theory
  of polarons and bipolarons in the half-filled holstein model},}\ }\href
  {\doibase 10.1103/PhysRevB.74.045106} {\bibfield  {journal} {\bibinfo
  {journal} {Phys. Rev. B}\ }\textbf {\bibinfo {volume} {74}},\ \bibinfo
  {pages} {045106} (\bibinfo {year} {2006})}\BibitemShut {NoStop}%
\bibitem [{\citenamefont {Yun}\ \emph {et~al.}(2007)\citenamefont {Yun},
  \citenamefont {Kang}, \citenamefont {Jeon},\ and\ \citenamefont
  {Choi}}]{yun_2007}%
  \BibitemOpen
  \bibfield  {author} {\bibinfo {author} {\bibfnamefont {Jae~Hyun}\
  \bibnamefont {Yun}}, \bibinfo {author} {\bibfnamefont {Hee~Soo}\ \bibnamefont
  {Kang}}, \bibinfo {author} {\bibfnamefont {Gun~Sang}\ \bibnamefont {Jeon}}, \
  and\ \bibinfo {author} {\bibfnamefont {Han-Yong}\ \bibnamefont {Choi}},\
  }\bibfield  {title} {\enquote {\bibinfo {title} {Dynamical mean-field theory
  of holstein model at half filling: Phonon frequency dependence of
  metal??insulator transition},}\ }\href {\doibase
  https://doi.org/10.1016/j.jmmm.2006.10.880} {\bibfield  {journal} {\bibinfo
  {journal} {Journal of Magnetism and Magnetic Materials}\ }\textbf {\bibinfo
  {volume} {310}},\ \bibinfo {pages} {916 -- 918} (\bibinfo {year}
  {2007})}\BibitemShut {NoStop}%
\bibitem [{\citenamefont {Barone}\ \emph {et~al.}(2008)\citenamefont {Barone},
  \citenamefont {Raimondi}, \citenamefont {Capone}, \citenamefont
  {Castellani},\ and\ \citenamefont {Fabrizio}}]{barone_2008}%
  \BibitemOpen
  \bibfield  {author} {\bibinfo {author} {\bibfnamefont {P.}~\bibnamefont
  {Barone}}, \bibinfo {author} {\bibfnamefont {R.}~\bibnamefont {Raimondi}},
  \bibinfo {author} {\bibfnamefont {M.}~\bibnamefont {Capone}}, \bibinfo
  {author} {\bibfnamefont {C.}~\bibnamefont {Castellani}}, \ and\ \bibinfo
  {author} {\bibfnamefont {M.}~\bibnamefont {Fabrizio}},\ }\bibfield  {title}
  {\enquote {\bibinfo {title} {Gutzwiller scheme for electrons and phonons: The
  half-filled hubbard-holstein model},}\ }\href {\doibase
  10.1103/PhysRevB.77.235115} {\bibfield  {journal} {\bibinfo  {journal} {Phys.
  Rev. B}\ }\textbf {\bibinfo {volume} {77}},\ \bibinfo {pages} {235115}
  (\bibinfo {year} {2008})}\BibitemShut {NoStop}%
\bibitem [{\citenamefont {Hecht}\ \emph {et~al.}(2008)\citenamefont {Hecht},
  \citenamefont {Weichselbaum}, \citenamefont {von Delft},\ and\ \citenamefont
  {Bulla}}]{hecht_2008}%
  \BibitemOpen
  \bibfield  {author} {\bibinfo {author} {\bibfnamefont {T}~\bibnamefont
  {Hecht}}, \bibinfo {author} {\bibfnamefont {A}~\bibnamefont {Weichselbaum}},
  \bibinfo {author} {\bibfnamefont {J}~\bibnamefont {von Delft}}, \ and\
  \bibinfo {author} {\bibfnamefont {R}~\bibnamefont {Bulla}},\ }\bibfield
  {title} {\enquote {\bibinfo {title} {Numerical renormalization group
  calculation of near-gap peaks in spectral functions of the anderson model
  with superconducting leads},}\ }\href
  {http://stacks.iop.org/0953-8984/20/i=27/a=275213} {\bibfield  {journal}
  {\bibinfo  {journal} {Journal of Physics: Condensed Matter}\ }\textbf
  {\bibinfo {volume} {20}},\ \bibinfo {pages} {275213} (\bibinfo {year}
  {2008})}\BibitemShut {NoStop}%
\bibitem [{\citenamefont {Toschi}\ \emph {et~al.}(2005)\citenamefont {Toschi},
  \citenamefont {Barone}, \citenamefont {Capone},\ and\ \citenamefont
  {Castellani}}]{toschi_2005b}%
  \BibitemOpen
  \bibfield  {author} {\bibinfo {author} {\bibfnamefont {A}~\bibnamefont
  {Toschi}}, \bibinfo {author} {\bibfnamefont {P}~\bibnamefont {Barone}},
  \bibinfo {author} {\bibfnamefont {M}~\bibnamefont {Capone}}, \ and\ \bibinfo
  {author} {\bibfnamefont {C}~\bibnamefont {Castellani}},\ }\bibfield  {title}
  {\enquote {\bibinfo {title} {Pairing and superconductivity from weak to
  strong coupling in the attractive hubbard model},}\ }\href
  {http://stacks.iop.org/1367-2630/7/i=1/a=007} {\bibfield  {journal} {\bibinfo
   {journal} {New Journal of Physics}\ }\textbf {\bibinfo {volume} {7}},\
  \bibinfo {pages} {7} (\bibinfo {year} {2005})}\BibitemShut {NoStop}%
\bibitem [{\citenamefont {Kuleeva}\ \emph {et~al.}(2014)\citenamefont
  {Kuleeva}, \citenamefont {Kuchinskii},\ and\ \citenamefont
  {Sadovskii}}]{kuleeva_2014}%
  \BibitemOpen
  \bibfield  {author} {\bibinfo {author} {\bibfnamefont {N.~A.}\ \bibnamefont
  {Kuleeva}}, \bibinfo {author} {\bibfnamefont {E.~Z.}\ \bibnamefont
  {Kuchinskii}}, \ and\ \bibinfo {author} {\bibfnamefont {M.~V.}\ \bibnamefont
  {Sadovskii}},\ }\bibfield  {title} {\enquote {\bibinfo {title} {Normal phase
  and superconducting instability in the attractive hubbard model: A dmft(nrg)
  study},}\ }\href {http://www.jetp.ac.ru/cgi-bin/e/index/r/146/2/p304?a=list}
  {\bibfield  {journal} {\bibinfo  {journal} {Journal of Experimental and
  Theoretical Physics}\ }\textbf {\bibinfo {volume} {119}},\ \bibinfo {pages}
  {264} (\bibinfo {year} {2014})}\BibitemShut {NoStop}%
\bibitem [{\citenamefont {Micnas}\ \emph {et~al.}(1990)\citenamefont {Micnas},
  \citenamefont {Ranninger},\ and\ \citenamefont
  {Robaszkiewicz}}]{ranninger_1990}%
  \BibitemOpen
  \bibfield  {author} {\bibinfo {author} {\bibfnamefont {R.}~\bibnamefont
  {Micnas}}, \bibinfo {author} {\bibfnamefont {J.}~\bibnamefont {Ranninger}}, \
  and\ \bibinfo {author} {\bibfnamefont {S.}~\bibnamefont {Robaszkiewicz}},\
  }\bibfield  {title} {\enquote {\bibinfo {title} {Superconductivity in
  narrow-band systems with local nonretarded attractive interactions},}\ }\href
  {\doibase 10.1103/RevModPhys.62.113} {\bibfield  {journal} {\bibinfo
  {journal} {Rev. Mod. Phys.}\ }\textbf {\bibinfo {volume} {62}},\ \bibinfo
  {pages} {113--171} (\bibinfo {year} {1990})}\BibitemShut {NoStop}%
\bibitem [{\citenamefont {Sandvik}\ \emph {et~al.}(2004)\citenamefont
  {Sandvik}, \citenamefont {Scalapino},\ and\ \citenamefont
  {Bickers}}]{sandvik_2004}%
  \BibitemOpen
  \bibfield  {author} {\bibinfo {author} {\bibfnamefont {A.~W.}\ \bibnamefont
  {Sandvik}}, \bibinfo {author} {\bibfnamefont {D.~J.}\ \bibnamefont
  {Scalapino}}, \ and\ \bibinfo {author} {\bibfnamefont {N.~E.}\ \bibnamefont
  {Bickers}},\ }\bibfield  {title} {\enquote {\bibinfo {title} {Effect of an
  electron-phonon interaction on the one-electron spectral weight of a d-wave
  superconductor},}\ }\href {\doibase 10.1103/PhysRevB.69.094523} {\bibfield
  {journal} {\bibinfo  {journal} {Phys. Rev. B}\ }\textbf {\bibinfo {volume}
  {69}},\ \bibinfo {pages} {094523} (\bibinfo {year} {2004})}\BibitemShut
  {NoStop}%
\bibitem [{\citenamefont {Bulla}\ \emph {et~al.}(2001)\citenamefont {Bulla},
  \citenamefont {Costi},\ and\ \citenamefont {Vollhardt}}]{bulla_2001}%
  \BibitemOpen
  \bibfield  {author} {\bibinfo {author} {\bibfnamefont {R.}~\bibnamefont
  {Bulla}}, \bibinfo {author} {\bibfnamefont {T.~A.}\ \bibnamefont {Costi}}, \
  and\ \bibinfo {author} {\bibfnamefont {D.}~\bibnamefont {Vollhardt}},\
  }\bibfield  {title} {\enquote {\bibinfo {title} {Finite-temperature numerical
  renormalization group study of the mott transition},}\ }\href {\doibase
  10.1103/PhysRevB.64.045103} {\bibfield  {journal} {\bibinfo  {journal} {Phys.
  Rev. B}\ }\textbf {\bibinfo {volume} {64}},\ \bibinfo {pages} {045103}
  (\bibinfo {year} {2001})}\BibitemShut {NoStop}%
\end{thebibliography}%


\begin{thebibliography}{7}%
\makeatletter
\providecommand \@ifxundefined [1]{%
 \@ifx{#1\undefined}
}%
\providecommand \@ifnum [1]{%
 \ifnum #1\expandafter \@firstoftwo
 \else \expandafter \@secondoftwo
 \fi
}%
\providecommand \@ifx [1]{%
 \ifx #1\expandafter \@firstoftwo
 \else \expandafter \@secondoftwo
 \fi
}%
\providecommand \natexlab [1]{#1}%
\providecommand \enquote  [1]{``#1''}%
\providecommand \bibnamefont  [1]{#1}%
\providecommand \bibfnamefont [1]{#1}%
\providecommand \citenamefont [1]{#1}%
\providecommand \href@noop [0]{\@secondoftwo}%
\providecommand \href [0]{\begingroup \@sanitize@url \@href}%
\providecommand \@href[1]{\@@startlink{#1}\@@href}%
\providecommand \@@href[1]{\endgroup#1\@@endlink}%
\providecommand \@sanitize@url [0]{\catcode `\\12\catcode `\$12\catcode
  `\&12\catcode `\#12\catcode `\^12\catcode `\_12\catcode `\%12\relax}%
\providecommand \@@startlink[1]{}%
\providecommand \@@endlink[0]{}%
\providecommand \url  [0]{\begingroup\@sanitize@url \@url }%
\providecommand \@url [1]{\endgroup\@href {#1}{\urlprefix }}%
\providecommand \urlprefix  [0]{URL }%
\providecommand \Eprint [0]{\href }%
\providecommand \doibase [0]{http://dx.doi.org/}%
\providecommand \selectlanguage [0]{\@gobble}%
\providecommand \bibinfo  [0]{\@secondoftwo}%
\providecommand \bibfield  [0]{\@secondoftwo}%
\providecommand \translation [1]{[#1]}%
\providecommand \BibitemOpen [0]{}%
\providecommand \bibitemStop [0]{}%
\providecommand \bibitemNoStop [0]{.\EOS\space}%
\providecommand \EOS [0]{\spacefactor3000\relax}%
\providecommand \BibitemShut  [1]{\csname bibitem#1\endcsname}%
\let\auto@bib@innerbib\@empty
\bibitem [{\citenamefont {Hofstetter}(2000)}]{hofstetter_2000}%
  \BibitemOpen
  \bibfield  {author} {\bibinfo {author} {\bibfnamefont {Walter}\ \bibnamefont
  {Hofstetter}},\ }\bibfield  {title} {\enquote {\bibinfo {title} {Generalized
  numerical renormalization group for dynamical quantities},}\ }\href {\doibase
  10.1103/PhysRevLett.85.1508} {\bibfield  {journal} {\bibinfo  {journal}
  {Phys. Rev. Lett.}\ }\textbf {\bibinfo {volume} {85}},\ \bibinfo {pages}
  {1508--1511} (\bibinfo {year} {2000})}\BibitemShut {NoStop}%
\bibitem [{\citenamefont {Bauer}\ \emph {et~al.}(2009)\citenamefont {Bauer},
  \citenamefont {Hewson},\ and\ \citenamefont {Dupuis}}]{bauer_2009}%
  \BibitemOpen
  \bibfield  {author} {\bibinfo {author} {\bibfnamefont {J.}~\bibnamefont
  {Bauer}}, \bibinfo {author} {\bibfnamefont {A.~C.}\ \bibnamefont {Hewson}}, \
  and\ \bibinfo {author} {\bibfnamefont {N.}~\bibnamefont {Dupuis}},\
  }\bibfield  {title} {\enquote {\bibinfo {title} {Dynamical mean-field theory
  and numerical renormalization group study of superconductivity in the
  attractive hubbard model},}\ }\href {\doibase 10.1103/PhysRevB.79.214518}
  {\bibfield  {journal} {\bibinfo  {journal} {Phys. Rev. B}\ }\textbf {\bibinfo
  {volume} {79}},\ \bibinfo {pages} {214518} (\bibinfo {year}
  {2009})}\BibitemShut {NoStop}%
\bibitem [{\citenamefont {Jeon}\ \emph {et~al.}(2003)\citenamefont {Jeon},
  \citenamefont {Park},\ and\ \citenamefont {Choi}}]{jeon_2003}%
  \BibitemOpen
  \bibfield  {author} {\bibinfo {author} {\bibfnamefont {Gun~Sang}\
  \bibnamefont {Jeon}}, \bibinfo {author} {\bibfnamefont {Tae-Ho}\ \bibnamefont
  {Park}}, \ and\ \bibinfo {author} {\bibfnamefont {Han-Yong}\ \bibnamefont
  {Choi}},\ }\bibfield  {title} {\enquote {\bibinfo {title} {Numerical
  renormalization-group study of the symmetric anderson-holstein model: Phonon
  and electron spectral functions},}\ }\href {\doibase
  10.1103/PhysRevB.68.045106} {\bibfield  {journal} {\bibinfo  {journal} {Phys.
  Rev. B}\ }\textbf {\bibinfo {volume} {68}},\ \bibinfo {pages} {045106}
  (\bibinfo {year} {2003})}\BibitemShut {NoStop}%
\bibitem [{\citenamefont {Hecht}\ \emph {et~al.}(2008)\citenamefont {Hecht},
  \citenamefont {Weichselbaum}, \citenamefont {von Delft},\ and\ \citenamefont
  {Bulla}}]{hecht_2008}%
  \BibitemOpen
  \bibfield  {author} {\bibinfo {author} {\bibfnamefont {T}~\bibnamefont
  {Hecht}}, \bibinfo {author} {\bibfnamefont {A}~\bibnamefont {Weichselbaum}},
  \bibinfo {author} {\bibfnamefont {J}~\bibnamefont {von Delft}}, \ and\
  \bibinfo {author} {\bibfnamefont {R}~\bibnamefont {Bulla}},\ }\bibfield
  {title} {\enquote {\bibinfo {title} {Numerical renormalization group
  calculation of near-gap peaks in spectral functions of the anderson model
  with superconducting leads},}\ }\href
  {http://stacks.iop.org/0953-8984/20/i=27/a=275213} {\bibfield  {journal}
  {\bibinfo  {journal} {Journal of Physics: Condensed Matter}\ }\textbf
  {\bibinfo {volume} {20}},\ \bibinfo {pages} {275213} (\bibinfo {year}
  {2008})}\BibitemShut {NoStop}%
\bibitem [{\citenamefont {Kuleeva}\ \emph {et~al.}(2014)\citenamefont
  {Kuleeva}, \citenamefont {Kuchinskii},\ and\ \citenamefont
  {Sadovskii}}]{kuleeva_2014}%
  \BibitemOpen
  \bibfield  {author} {\bibinfo {author} {\bibfnamefont {N.~A.}\ \bibnamefont
  {Kuleeva}}, \bibinfo {author} {\bibfnamefont {E.~Z.}\ \bibnamefont
  {Kuchinskii}}, \ and\ \bibinfo {author} {\bibfnamefont {M.~V.}\ \bibnamefont
  {Sadovskii}},\ }\bibfield  {title} {\enquote {\bibinfo {title} {Normal phase
  and superconducting instability in the attractive hubbard model: A dmft(nrg)
  study},}\ }\href {http://www.jetp.ac.ru/cgi-bin/e/index/r/146/2/p304?a=list}
  {\bibfield  {journal} {\bibinfo  {journal} {Journal of Experimental and
  Theoretical Physics}\ }\textbf {\bibinfo {volume} {119}},\ \bibinfo {pages}
  {264} (\bibinfo {year} {2014})}\BibitemShut {NoStop}%
\bibitem [{\citenamefont {Wilson}(1975)}]{wilson_1975}%
  \BibitemOpen
  \bibfield  {author} {\bibinfo {author} {\bibfnamefont {Kenneth~G.}\
  \bibnamefont {Wilson}},\ }\bibfield  {title} {\enquote {\bibinfo {title} {The
  renormalization group: Critical phenomena and the kondo problem},}\ }\href
  {\doibase 10.1103/RevModPhys.47.773} {\bibfield  {journal} {\bibinfo
  {journal} {Rev. Mod. Phys.}\ }\textbf {\bibinfo {volume} {47}},\ \bibinfo
  {pages} {773--840} (\bibinfo {year} {1975})}\BibitemShut {NoStop}%
\bibitem [{\citenamefont {Krishna-murthy}\ \emph {et~al.}(1980)\citenamefont
  {Krishna-murthy}, \citenamefont {Wilkins},\ and\ \citenamefont
  {Wilson}}]{krishnamurthy_1980}%
  \BibitemOpen
  \bibfield  {author} {\bibinfo {author} {\bibfnamefont {H.~R.}\ \bibnamefont
  {Krishna-murthy}}, \bibinfo {author} {\bibfnamefont {J.~W.}\ \bibnamefont
  {Wilkins}}, \ and\ \bibinfo {author} {\bibfnamefont {K.~G.}\ \bibnamefont
  {Wilson}},\ }\bibfield  {title} {\enquote {\bibinfo {title}
  {Renormalization-group approach to the anderson model of dilute magnetic
  alloys. i. static properties for the symmetric case},}\ }\href {\doibase
  10.1103/PhysRevB.21.1003} {\bibfield  {journal} {\bibinfo  {journal} {Phys.
  Rev. B}\ }\textbf {\bibinfo {volume} {21}},\ \bibinfo {pages} {1003--1043}
  (\bibinfo {year} {1980})}\BibitemShut {NoStop}%
\end{thebibliography}%
\end{document}


\title{Supplemental Material for\\Dynamical effects on superconductivity in BCS-BEC crossover}

\author{Tae-Ho Park}
\affiliation{Department of Physics and Institute for Basic Science Research, Sungkyunkwan University, Suwon 16419,
Korea}

\author{Han-Yong Choi}
\affiliation{Department of Physics and Institute for Basic Science Research, Sungkyunkwan University, Suwon 16419,
Korea}
\affiliation{Asia Pacific Center for Theoretical Physics, Pohang 37673, Korea}

\date{\today}

\begin{abstract}

\end{abstract}

\pacs{}

\maketitle

\onecolumngrid

\section{NRG-DMFT calculation}\label{sec:model}

The superconducting pairing term breaking the particle number symmetry is implemented in the bath of the effective impurity Hamiltonian
\begin{subequations}
\begin{equation}\label{eq:HolSu}
{\cal H}_{Hol}^{SC} = \sum_{k,\sigma} \varepsilon_k c_{k\sigma}^\dag c_{k\sigma}
+ \sum_{k,\sigma} V_k \left( f_{\sigma}^\dag c_{k\sigma} +
c_{k\sigma}^\dag f_{\sigma} \right) \\
 -\sum_{k,\sigma} \Delta_k^{SC} \left( c_{k\uparrow}^\dag c_{-k\downarrow}^\dag
+c_{-k\downarrow} c_{k\uparrow} \right)  + {\cal H}_{loc},
\end{equation}
\begin{equation}\label{eq:loc}
{\cal H}_{loc} =  \omega_0  a^\dag a - \mu n_f + g ( a^\dag +a) ( n_f -1 ),
\end{equation}
\end{subequations}
where $f_\sigma$ is the fermionic operator on the impurity site, $n_f=\sum_\sigma f_\sigma^\dag f_\sigma$, and we solve Eq.\ (\ref{eq:HolSu}) using density matrix numerical renormalization group (DM--NRG) technique \cite{hofstetter_2000}. $\varepsilon_k$, $V_k$, and
$\Delta_k^{SC}$ are parameters of the effective medium for the DMFT self-consistent
calculation and $\mu$ is the chemical potential. We follow the Lanczos transformation for the NRG  calculation from Ref. \onlinecite{bauer_2009}.
We take the NRG parameters $\Lambda=1.6$, $N_S=1000$, $N_{ph}=40$ for numerical calculations.

For nonzero $\Delta_k^{SC}$, the Hamiltonian of Eq.\ (\ref{eq:HolSu}) introduces not only
a single-particle Green's function $G_k(\omega)=\llangle c_{k,\uparrow}, c_{k\uparrow}^\dag \rrangle_\omega$
but also an anomalous Green's function $F_k(\omega)=\llangle c_{k,\uparrow}, c_{-k\downarrow} \rrangle_\omega$.
The double bracket $\llangle ~~ \rrangle$ represents the correlator which is defined as in Ref.\ \onlinecite{jeon_2003} as
 \ba
\llangle {\cal O}_1,{\cal O}_2 \rrangle_\omega &=&
\int_{-\infty}^{\infty} dt\ e^{i\omega t}\ \llangle {\cal
O}_1,{\cal O}_2 \rrangle_t ,
 \\ \nonumber
\llangle {\cal O}_1,{\cal O}_2 \rrangle_t &=& -i \theta(t) \left<
[{\cal O}_1 (t),{\cal O}_2 (0)]_\zeta \right>=-i \theta(t) \left<
[{\cal O}_1 (0),{\cal O}_2 (-t)]_\zeta \right>.
 \ea
Here, $\theta$ is the step function, $\left<~~\right>$ the
thermodynamic average, and $\zeta = +$ if both ${\cal O}_1$ and
${\cal O}_2$ are fermion operators and $\zeta = -$ otherwise. The
equation of motion of the correlator is given by
 \ba
 \label{eom}
\omega \llangle {\cal O}_1,{\cal O}_2 \rrangle_\omega &=& \left< [
{\cal O}_1,{\cal O}_2 ]_\zeta \right> + \llangle [{\cal
O}_1,{\cal H} ]_- , {\cal O}_2 \rrangle_\omega ,
 \\ \nonumber
&=& \left< [ {\cal O}_1,{\cal O}_2 ]_\zeta \right> - \llangle
{\cal O}_1 , [{\cal O}_2,{\cal H} ]_- \rrangle_\omega .
 \ea

To describe the superconductivity, it is convenient to use the Nambu spinor representation as
$\mathbf{\Psi}_k^\dag \equiv \left( c_{k,\uparrow}^\dag, c_{-k,\downarrow}\right)$,
then the lattice Green's function can be represented by matrix form as
\begin{equation}\label{eq:GFM}
\hat{\mathbf{G}}_k(\omega)=
\left( \begin{array}{cc}
G_k(\omega)  & F_k(\omega) \\
F_k(- \omega)^\ast & -G_{-k}(-\omega)^\ast
\end{array} \right),
\end{equation}
and using the Dyson equation in terms of the local self-energy $\hat{\mathbf{\Sigma}}( \omega )$
for infinite dimensions, the Green's function Eq.\ (\ref{eq:GFM}) is expressed as
$\hat{\mathbf{G}}_k(\omega)^{-1}=\omega \tau_0-(\varepsilon_k-\mu)  \tau_3-\hat{\mathbf{\Sigma}}(\omega)$,
where $\tau_i$ ($i=1,2,3$) is the Pauli matrix, $\tau_0$ is the $2 \times 2$
identity matrix, and $\mu$ is the chemical potential. In the Holstein model, the self-energy is attributed to
the electron-phonon coupling term and the relevant correlation functions from the effective impurity model are
\begin{equation}\label{eq:EPC}
\hat{\mathbf{C}}(\omega)=
\left( \begin{array}{cc}
\llangle (a^{\dag}+a) f_{\uparrow}, f_{\uparrow}^\dag \rrangle_\omega  & \llangle (a^{\dag}+a) f_{\uparrow}, f_{\downarrow} \rrangle_\omega \\
-\llangle (a^{\dag}+a) f_{\downarrow}^\dag, f_{\uparrow}^\dag \rrangle_\omega  & -\llangle (a^{\dag}+a) f_{\downarrow}^\dag, f_{\downarrow} \rrangle_\omega
\end{array} \right),
\end{equation}
and the self-energy is obtained by $\hat{\mathbf{\Sigma}}( \omega )=g \hat{\mathbf{C}}(\omega) \hat{\mathbf{G}}_f(\omega)^{-1}$ where
\begin{equation}\label{eq:Gf}
\hat{\mathbf{G}}_f(\omega)=
\left( \begin{array}{cc}
\llangle f_{\uparrow}, f_{\uparrow}^\dag \rrangle_\omega  & \llangle  f_{\uparrow}, f_{\downarrow} \rrangle_\omega \\
\llangle f_{\downarrow}^\dag, f_{\uparrow}^\dag \rrangle_\omega  & \llangle  f_{\downarrow}^\dag, f_{\downarrow} \rrangle_\omega
\end{array} \right).
\end{equation}
For the matrix convenience, we decompose the self-energy matrix to
$\hat{\mathbf{\Sigma}}( \omega )=\Sigma (\omega) \tau_0+\phi (\omega) \tau_1$
and  define a function $W(\omega) \equiv \omega-\Sigma (\omega)$ for the diagonal component.
Then, the lattice Green's function can be simply rewritten as
\begin{equation}\label{eq:GFSC}
\hat{\mathbf{G}}_k(\omega)^{-1}=W(\omega)\tau_0 - (\varepsilon_k-\mu)  \tau_3 - \phi(\omega) \tau_1.
\end{equation}

The self-consistency for DMFT calculation is achieved by using the local Green's
function $\hat{\mathbf{G}}^{loc}(\omega)$ ,
\begin{equation}\label{eq:selfcon}
  \hat{\cal G}_{0}^{-1}(\omega)= \hat{\mathbf{G}}^{loc}(\omega)^{-1}
  +\hat{\mathbf{\Sigma}}( \omega ),
\end{equation}
where $\hat{\cal G}_{0}^{-1}(\omega)$ is matrix form of the Weiss field
and the local Green's function is obtained by momentum ($k$) summation
($ \hat{\mathbf{G}}^{loc}(\omega)= \sum_k \hat{\mathbf{G}}_k(\omega)$).
By considering the semi-elliptic bare density of state per spin $D(\epsilon)=(2/\pi)\sqrt{D^2-\epsilon^2}/D^2 $
for the infinite dimensional Bethe lattice, the local Green's function can be obtained as
\begin{equation}\label{eq:Gloc}
\begin{split}
\hat{\mathbf{G}}^{loc}(\omega)= \int d\epsilon D(\epsilon) \left[ W(\omega)\tau_0 - \epsilon  \tau_3 - \phi(\omega) \tau_1 \right]^{-1}\\
= \frac{2}{D^2} \left[ 1-s \frac{\sqrt{W(\epsilon)^2-\phi(\epsilon)^2-D^2}}{\sqrt{W(\epsilon)^2-\phi(\epsilon)^2}} \right] \left[  W (\omega) \tau_0+\phi (\omega) \tau_1 \right],
\end{split}
\end{equation}
where $s=\textrm{sgn}\left[ \textrm{Im} \left( \sqrt{W(\epsilon)^2-\phi(\epsilon)^2} \right) \right]$,
and the Weiss field is obtained as
\begin{equation}\label{eq:Weiss}
  \hat{\cal G}_{0}^{-1}(\omega)= \omega \tau_0 +\mu \tau_3- \frac{D^2}{4} \tau_3 \hat{\mathbf{G}}^{loc}(\omega) \tau_3.
\end{equation}

\begin{figure*}
\begin{center}
\includegraphics[scale=0.8]{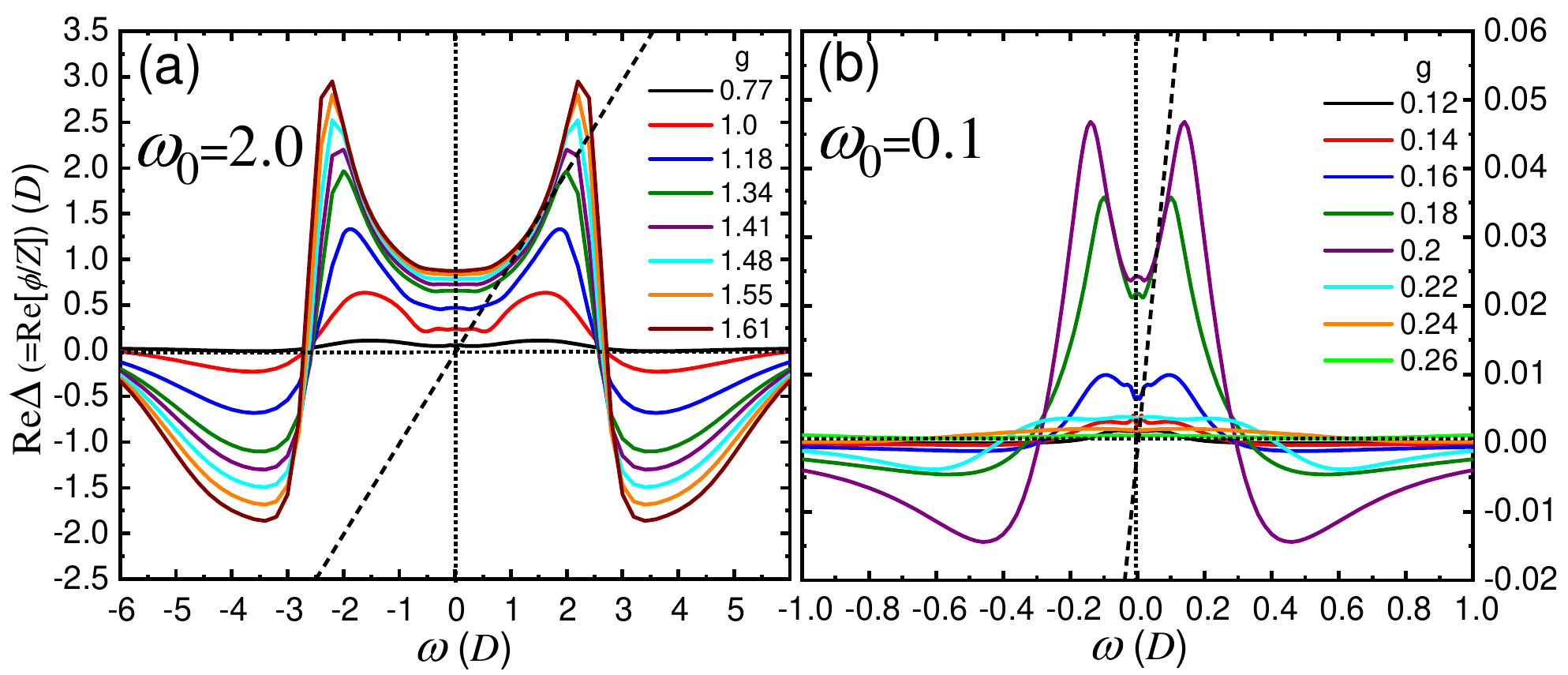}
\caption{ The real part of the gap functions for (a) anti-adiabatic and (b) adiabatic cases of
Holstein model. The black dashed lines are the $y=\omega$ linear line and
the intersection points correspond to the pairing gap $\Delta_p$.
}
\label{fig:ReGap}
\end{center}
\end{figure*}

\section{Pairing gap, superfluid stiffness and critical temperature}\label{sec:resultsA}

For the Holstein model of Eq.\ (1) of the main text, the chemical potential is equal to zero at half-filling because the band is particle-hole symmetric about $\mu=0$. We get
the gap function by renormalization of the off-diagonal self-energy as
$\Delta(\omega)=\phi(\omega)/Z(\omega)$,
where the renormalization function is $Z(\omega)=1-\Sigma(\omega)/\omega$.
Then, the pairing gap $ \Delta_p$ is given by the solution of the equation
$ \Delta_p= \textrm{Re}\Delta(\omega = \Delta_p)$. This procedure is illustrated in Fig.\ 1.
The solutions are graphically the intersection points of $\textrm{Re} \Delta (\omega)$
and the linear line ($y=\omega $). Figure (a) is for the anti-adiabatic case of $\omega_0=2.0 D$ and (b) is for the adiabatic case of $\omega_0= 0.1 D$.
The behaviors of $\Delta_p$ as a function of the electron-phonon coupling $g$ are different depending on the phonon frequency $\omega_0$ as shown in Fig.\ 2 in the main text. $\Delta_p$ increases monotonically as a function of $g$ for anti-adiabatic cases like the attractive-Hubbard model.
For adiabatic cases, however, $\Delta_p$ increases initially and gets suppressed in BEC regime due to the band narrowing polaronic effects.
Also, for anti-adiabatic cases, multiple solutions exist for $g/D>1.34$ as shown in Fig.\ 1(a),
which may cause complicated structures in physical properties like density of states.

The band narrowing by the adiabatic phonons also affect the superfluid stiffness $D_S$. It decreases more rapidly in adiabatic cases than in anti-adiabatic cases as $g$ increases as presented in Fig.\ 2(a) and (b). $D_S$ determines the phase coherence of the SC phase and, subsequently, the coherence peak of density of states as was presented in Fig.\ 3 in the main text.
$D_S$ is calculated from the zero
frequency weight of the real part of optical conductivity.
\begin{equation}\label{eq:SFstiffness}
D_S=-\frac{8}{\pi} \int d\varepsilon_k D(\varepsilon_k)V(\varepsilon_k) \int_{-\infty}^0 d\omega \textrm{Im}G_k^{\textrm{off}}(\omega) \textrm{Re}G_k^{\textrm{off}}(\omega),
\end{equation}
where $V(\epsilon)=(D^2-\epsilon^2)/3$
for the Bethe lattice and
\begin{equation}\label{eq:Goff}
 G^{\textrm{off}}(\epsilon,\omega)
=\frac{\phi(\omega)}{W(\omega)^2-\epsilon^2-\phi(\omega)^2}.
\end{equation}
By inserting Eq.\ (\ref{eq:Goff}) to Eq.\ (\ref{eq:SFstiffness}), we obtain
\begin{equation}\label{eq:stiffness}
D_S=-\frac{8}{3\pi D^2}  \int_{-\infty}^0 d\omega \left[ 1- \frac{\sqrt{a(\omega)^2-1}}{a(\omega)} -\frac{\sqrt{a(\omega)^2-1}}{2a(\omega)^3} \right],
\end{equation}
where $a(\omega)=\sqrt{W(\omega)^2-\phi(\omega)^2}/D $.

\begin{figure}
\begin{center}
\includegraphics[scale=0.75]{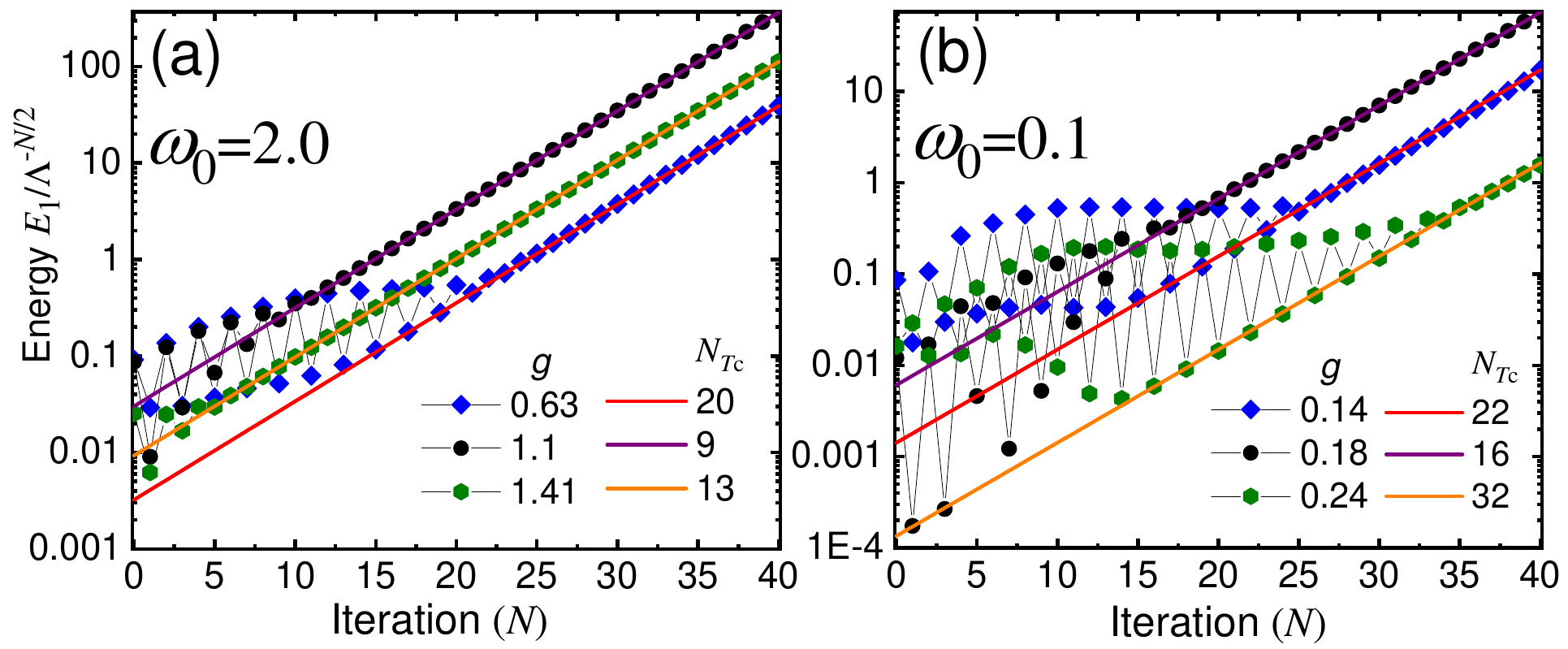}
 \caption{ The low lying excitation energy flows for (a) anti-adiabatic and (b) adiabatic cases. Three representative cases of the weak, intermediate, and strong electron-phonon coupling regimes are presented. The solid straight lines are the fitting function $\Lambda^{(N-N_{Tc})/2}$ from which the $T_c$ are determined.}
\label{fig:energyflow}
\end{center}
\end{figure}

The $T_c$ (black asterisks in Fig.\ \ref{fig:gap_stiffness})
is obtained from the exponentially diverging energy flow approaching the SC phase fixed point indicating an instability of the normal phase in the NRG-DMFT iteration procedure \cite{hecht_2008,kuleeva_2014}.
In the numerical renormalization group calculation, the renormalization group flow
can be presented by the variation of the NRG site number $N$.
As $N$ increases the energy scale and the temperature of the system goes down logarithmically.
Thus, the temperature can be changed by the variation of $N$ \cite{wilson_1975,krishnamurthy_1980}.
Therefore, the pairing energy scale can be obtained from the energy spectra ($E_N$)
and the corresponding energy flows determine the superconducting transition temperature $T_c$ \cite{hecht_2008}.
The NRG energy flows $E_N/\Lambda^{-N/2}$ are fitted using exponential function $\Lambda^{(N-N_{Tc})/2}$.

The superconducting critical temperatures $T_c$ can be determined by fitting the low lying
excitation energy flows $E_N/\Lambda^{-N/2}$ with respect to the NRG site number $N$ by $\Lambda^{(N-N_{Tc})/2}$ as shown in Fig.\ \ref{fig:energyflow} \cite{hecht_2008}.
In the higher temperature region with small $N$ before entering the superconducting state,
the energy eigenvalues deviate for even and odd number of $N$ with same total spin ($S$) of the eigenstates.
The deviation ends up by passing through the pairing energy scale $\sim \Delta_{P}$.
In the anti-adiabatic case, the pairing energy scale monotonically increases as the electron-phonon coupling
$g$ increases as shown in the Fig.\ \ref{fig:energyflow}(a).
However, Fig.\ \ref{fig:energyflow}(a)
shows that the pairing energy scale decrease for the BEC regime with large $g$ for adiabatic phonon case,
which is consistent with the Fig.\ 2 in the main manuscript.
Then, the $T_c$ is determined by
 \ba
 T_c = \Lambda^{-N_{Tc}/2}.
 \ea
We present the excitation energy flows in the logarithmic scale in Fig.\ \ref{fig:energyflow}(a) and (b)
for adiabatic and antiadiabatic cases, respectively, and $N_{Tc} \sim -2 \log y_0/\log \Lambda$,
where $y_0$ is the y-axis intercept. The calculated $T_c$ are shown in Fig.\ 2 of the main text.

\section{Self-energy in BEC regime}

\begin{figure}
\begin{center}
\includegraphics[scale=0.75]{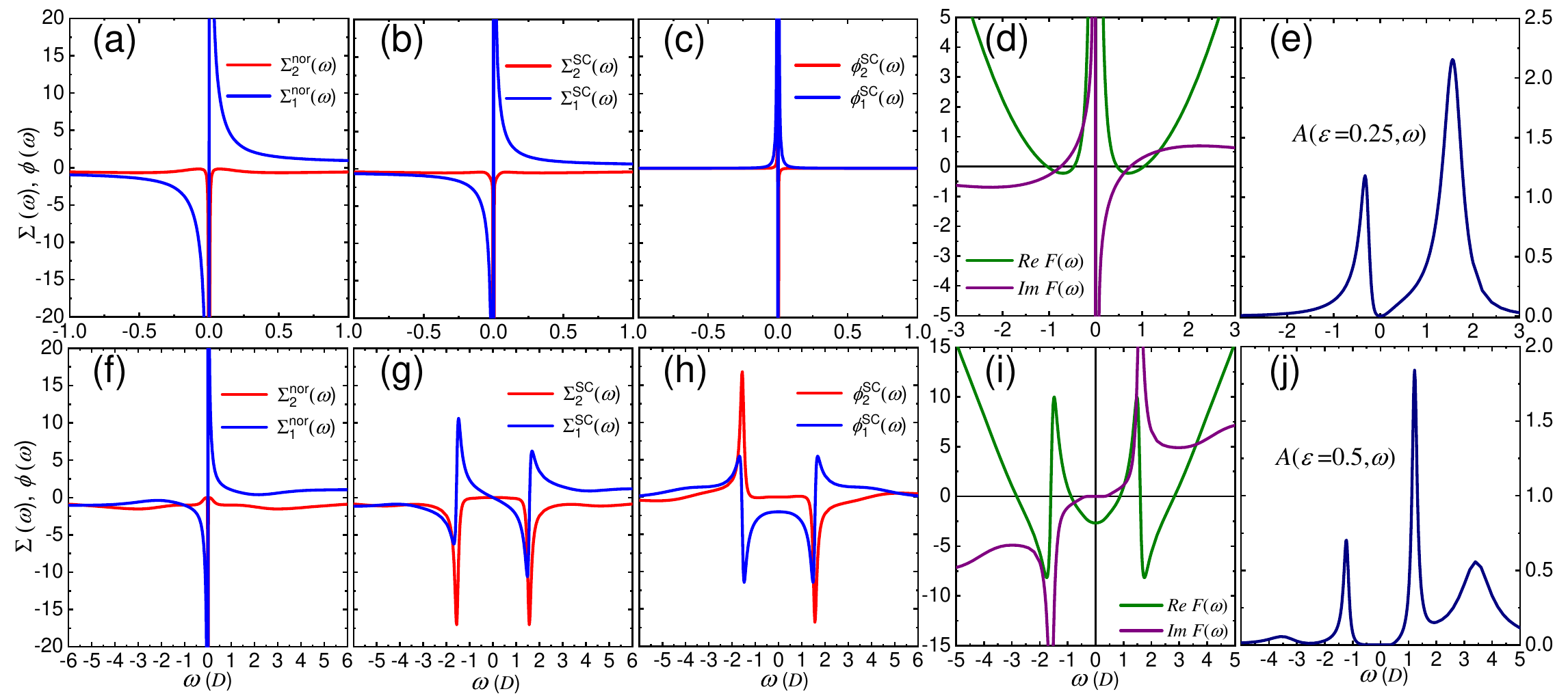}
 \caption{ The self-energies corresponding to the plots (c) and (f) of Fig.\ 4 in the main text. The upper row is for the adiabatic case of $\omega_0 = 0.1$ and the bottom row for the anti-adiabatic case of $\omega_0=2$. The plots in the 1st, 2nd, and 3rd columns are the diagonal self-energy in normal state, the diagonal self-energy in SC state, and the off-diagonal self-energy, respectively. In the 4th column we show the function $F(\omega)=W(\omega)^2-\phi(\omega)^2$. The intersection of $y=Re F(\omega)$ and $y=\epsilon^2$ determines the SC dispersion. In the 5th column we shown thus determined spectral function as a function of $\omega$. (e) is at $\epsilon=0.25$ for adiabatic case and (j) is at $\epsilon=0.5$ for anti-adiabatic case.
 }
\label{fig:sup3}
\end{center}
\end{figure}

We show in Fig.\ \ref{fig:sup3} the self-energies and related plots corresponding to the plots (c) and (f) in Fig.\ 4 of the main text. The upper row is for the adiabatic case of $\omega_0 = 0.1$ and the bottom row for the anti-adiabatic case of $\omega_0=2$. The plots in the 1st, 2nd, and 3rd columns are the diagonal self-energy in normal state, the diagonal self-energy in SC state, and the off-diagonal self-energy, respectively. In plots (d) and (i) we show the function $F(\omega)\equiv W(\omega)^2-\phi(\omega)^2$, where $W(\omega) =\omega -\Sigma(\omega)$. The subscripts 1 and 2, respectively, stand for the real and imaginary parts.
The intersection of $y=Re F(\omega)$ and $y=\epsilon^2$ determines the quasi-particle-dispersion in SC state.

Consider the anti-adiabatic case first. For $\epsilon=0$ there are three intersection points for $\omega>0$ region. Label them as $\omega_1$, $\omega_2$, and $\omega_3$. ($\omega_1<\omega_2<\omega_3$). The spectral function is given by
 \ba
 A(\epsilon,\omega) = -\frac1\pi Im \frac{W(\omega) +\epsilon}{W^2 -\phi^2 -\epsilon^2}.
 \ea
Note from the plot (i) that the imaginary part of the function $F(\omega)$ is such that $Im F(\omega_1) < Im F(\omega_3)<Im F(\omega_2)$. Therefore, the spectral function exhibits a peak near $\omega_1$ and a smaller peak around $\omega_3$, but a dip around $\omega \approx \omega_2$. The numerator of $W(\omega) +\epsilon$ is anti-symmetric with respect to $\omega$ for $\epsilon=0$ and the spectral function $A$ is symmetric.
As $\epsilon$ increases, the intersection point $\omega_1$ increases and $\omega_3$ increases more rapidly, as can be seen from the plot (i). The symmetricity with respect to $\omega=0$ does not hold for $\epsilon \ne 0$. The branch closer to the bare dispersion appears with a stronger weight. For $\epsilon = 0.5$ the resulting $A(\epsilon,\omega)$ vs.\ $\omega$ is shown in the plot (j).

For the adiabatic case, there are two intersection points $\omega_1$ and $\omega_2$ ($\omega_1<\omega_2$) for $\omega>0$ as can be seen from the plot (d). Because they are close only one peak appears for $\omega>0$ in the spectral function. As $\epsilon$ increases, the $\omega_1$ decreases and $\omega_2$ increases as can be seen from the plot (d).
$A(\epsilon,\omega)$ is not symmetric about $\omega=0$ for $\epsilon\ne 0$. In the spectral function, a peak around $\omega_2$ appears for $\omega>0$ and a peak around $\omega \approx -\omega_1$ appears with a smaller weight for $\omega<0$.
In the plot (e), we show $A(\epsilon,\omega)$ vs.\ $\omega$ for $\epsilon = 0.25$.
The peak position in the negative $\omega$ region ($\omega_1$) keeps decreasing as $\epsilon$ increases which means an absence of the back-bending in the quadrant of $\epsilon>0$ and $\omega<0$ of Fig.\ 4 of the main text. This is due to the increases of $Re[ W(\omega)^2-\phi(\omega)^2 ]$ as $\omega \rightarrow 0$, which can be traced back to the increases of $Re \Sigma(\omega) $ as $\omega$ approaches 0 from above. Because $Re \Sigma(\omega) $ is an odd function of $\omega$ this means that the slope of $Re \Sigma(\omega) \rightarrow \infty$ as $\omega \rightarrow 0$ as discussed in the main text.

\bibliography{reference.bib}{}